\documentclass[aps,twocolumn,showpacs,superscriptaddress,preprintnumbers,floatfix,amsmath,amssymb,eufrak,table]{revtex4}
\usepackage{ulem}
\usepackage{color}
\usepackage{graphicx}  
\usepackage{appendix}
 \usepackage{epsfig}
\usepackage[colorlinks=true,linkcolor= red, citecolor = cyan, urlcolor = blue ]{hyperref}
\usepackage{epstopdf}
\usepackage{amsmath}
\usepackage{subfig}
\usepackage{comment}
\usepackage{float}
\usepackage{multirow}
\usepackage{ragged2e} 

\newcommand{\be}{\begin{equation}}
	\newcommand{\ee}{\end{equation}}
\newcommand{\ba}{\begin{eqnarray}}
	\newcommand{\ea}{\end{eqnarray}}

\begin{document}
\title{Systematic study of bottomonium production in proton–proton collisions at LHC energies}
	
\author{Biswarup Paul}\email{biswarup.babu@gmail.com}
\affiliation{Bali Ram Bhagat College, Samastipur - 848101, Bihar, India}

\date{\today}
\begin{abstract}
We present a comprehensive study of $\Upsilon(nS)$ ($n$ = 1, 2, 3) production in proton–proton ($pp$) collisions at various LHC energies and rapidity ranges within the framework of leading order non-relativistic quantum chromodynamics (NRQCD) factorization. The transverse momentum ($p_{\rm T}$)–dependent production cross-sections are calculated, incorporating both direct and feed-down contributions. Specifically, the feed-down contributions to $\Upsilon(1S)$ include those
from $\Upsilon(2S)$, $\Upsilon(3S)$, $\chi_{bJ}(1P)$, and $\chi_{bJ}(2P)$,
while $\Upsilon(2S)$ receives feed-down from $\Upsilon(3S)$ and
$\chi_{bJ}(2P)$. For $\Upsilon(3S)$, feed-down from the
$\chi_{bJ}(3P)$ states is also taken into account. The computed cross-sections and cross-section ratios among different $\Upsilon$ states are compared with experimental measurements from ALICE, ATLAS, CMS and LHCb. It is found that the experimental cross-sections and cross-section ratios are well described within the theoretical uncertainties arising due to the choices of the factorization and renormalization scales for $p_{\rm T}$ $>$ 4 GeV and $p_{\rm T}$ $>$ 0 GeV, respectively. Furthermore, the cross-section ratios exhibit a clear saturation behavior beyond $p_{\rm T}$ $\approx$ 40 GeV.
\end{abstract}
\pacs{Quarkonium, NRQCD, proton--proton collisions, LHC}
	
\maketitle

\section{Introduction}

The study of heavy quarkonium production in high–energy hadronic collisions plays a central role in testing the dynamics of Quantum Chromodynamics (QCD). Because heavy quarks possess masses much larger than the QCD scale $\Lambda_{\mathrm{QCD}}$, the formation of bound states such as $J/\psi$ and $\Upsilon$ can be analyzed within a framework that combines perturbative and nonperturbative aspects of QCD. Among the theoretical approaches developed for this purpose, the non-relativistic quantum chromodynamics (NRQCD) factorization formalism provides the most widely used description of heavy quarkonium production and decay processes~\cite{prd51,my_jpg}.

Quarkonium production in NRQCD proceeds in two steps. First, the creation of a $Q\bar Q$ pair in a hard scattering process at short distances, is computed perturbatively as a series in $\alpha_s$, allowing the pair to be produced in either a color-singlet (CS) or color-octet (CO) configuration. Second, the nonperturbative evolution of the pair into a physical quarkonium state, is described by long-distance matrix elements (LDMEs), which encapsulate the soft-gluon emissions required to form a color-neutral bound state. The LDMEs are assumed to be universal and are determined through global fits to experimental data.


The color-octet contribution in the quarkonium wave function is a distinctive feature of the NRQCD approach~\cite{}. Before NRQCD, quarkonium production and decay were described using the Color-Singlet Model (CSM) \cite{CSM1,CSM2}, where the heavy quark pair $Q\bar{Q}$ was assumed to be produced directly as a color-singlet state and higher-order terms in the relative velocity $v$ ($v$ is the velocity of the heavy quark in the quarkonium rest frame) were neglected. Although the CSM successfully described low-energy $J/\psi$ data, it was incomplete due to inconsistencies for P-wave quarkonia arising from non-factorizing infrared divergences observed in $\chi_c$ decays \cite{chi_c_decay}. This problem was resolved by including color-octet contributions for P-wave states \cite{CO_P_Wave}.

The inclusion of color-octet components significantly improved the description of charmonium production at the Fermilab Tevatron \cite{pheno_Tevatron1,pheno_Tevatron2,pheno_Tevatron3}, particularly at high $p_{\rm T}$, where CSM predictions were inadequate \cite{CDF_JPsi_PsiP,CDF_JPsi_chi_c}. Although color-octet contributions are suppressed for S-wave states by $O(v^4)$, experimental measurements of direct $J/\psi$ and $\psi'$ production at the Tevatron indicate that these contributions are important \cite{CO_S_Wave}. Consequently, NRQCD provides a predictive framework for describing quarkonium physics. However, significant discrepancies remain in the description of the $p_{\rm T}$ dependence of quarkonium polarization in $J/\psi$ and $\Upsilon$ hadroproduction. Theoretical predictions within NRQCD at both leading order (LO) and next-to-leading order (NLO) fail to reproduce the observed polarization patterns. 


Over the past two decades, considerable progress has been achieved in NLO QCD calculations for quarkonium production. The NLO corrections to CS $J/\psi$ hadroproduction were studied in Refs.~\cite{NLO_CS_JPsi1,NLO_CS_JPsi2}, where it was shown that the $p_{\rm T}$ distribution is enhanced by two to three orders of magnitude in the high-$p_{\rm T}$ region. 
In Ref.~\cite{NLO_CS_JPsi2}, the NLO contributions to $J/\psi$ polarization are studied for the first time, revealing a strong shift from transverse polarization at LO to predominantly longitudinal polarization at NLO. However, the predicted longitudinal polarization remains larger than that observed in the Tevatron data~\cite{CDF_JPsi_PsiP_Pol_1p96}.
These features were later reproduced at LO within a new factorization framework designed for high-$p_{\rm T}$ quarkonium production~\cite{New_Model_LO}. Further studies investigated the NLO contributions to $J/\psi$ production through S-wave CO states (${}^{1}S_{0}^{[8]}, {}^{3}S_{1}^{[8]}$) in Ref.~\cite{NLO_CO_S_Wave}, where both the $p_{\rm T}$ distribution of the yield and the polarization were found to exhibit only small changes relative to the LO results. The NLO corrections for $\chi_{cJ}$ hadroproduction were analyzed in Ref.~\cite{NLO_chi_c}. Comprehensive calculations and global fits including both the yield and polarization of $J/\psi$ production at NLO QCD were later carried out by three independent groups~\cite{NLO_JPsi_Pol1,NLO_JPsi_Pol2,NLO_JPsi_Pol3}. In particular, Ref.~\cite{NLO_JPsi_Pol3} performed the first complete study of prompt $J/\psi$ and $\psi$(2S) hadroproduction, the results agree with CDF Run I data~\cite{CDF_JPsi_PsiP_Pol_1p8} but conflict with CDF Run II results~\cite{CDF_JPsi_PsiP_Pol_1p96}, and are consistent with ALICE measurements~\cite{ALICE_JPsi_PsiP_Pol}, though for inclusive $J/\psi$.
Despite these developments, the long-standing $J/\psi$ polarization puzzle remains unresolved. 

The LHCb Collaboration reported measurements of $\eta_c$ production~\cite{LHCb_eta_c}. This result triggered several theoretical investigations~\cite{NLO_chi_c1,NLO_chi_c2,NLO_chi_c3} that analyzed the data from different perspectives. In Ref.~\cite{NLO_chi_c1}, the $\eta_c$ measurement was interpreted as a potential challenge to the NRQCD framework, whereas Refs.~\cite{NLO_chi_c2,NLO_chi_c3} emphasized its implications for understanding $J/\psi$ production and polarization. The theoretical situation therefore remains complex and indicates that further phenomenological studies and tests of NRQCD are still necessary.

Bottomonium production and polarization provide an alternative and particularly clean environment for probing the hadronization of heavy quark pairs. Because the bottom quark mass is approximately three times larger than that of the charm quark, both the QCD coupling constant $\alpha_s~(\sqrt{4m_q^2 + p_T^2})$ and the velocity parameter $v^2$ are smaller. As a result, perturbative calculations based on the double expansion in $\alpha_s$ and $v^2$ are expected to converge more effectively for bottomonium than for charmonium. 
For $\Upsilon$ hadroproduction, similar theoretical progress has been achieved in the description of the $p_{\rm T}$ dependence of both the yield and polarization within the CS channel at QCD NLO~~\cite{NLO_CS_JPsi1,NLO_CS_JPsi2,NLO_CS_UPsi} and partially at next-to-next-to-leading order (NNLO) in Ref.~\cite{NNLO_CS_UPsi}. The NLO QCD corrections to the $p_{\rm T}$ dependence of the yield and polarization for $\Upsilon(1S, 3S)$ production through S-wave CO states (${}^{1}S_{0}^{[8]}, {}^{3}S_{1}^{[8]}$) have been presented in Ref.~\cite{NLO_CO_UPsi1}. In addition, the NLO QCD correction to the $p_{\rm T}$ distribution of the yield for $\Upsilon(1S)$ production via all CO states, including the ${}^{3}P_{J}^{[8]}$ channel, has been studied in Ref.~\cite{NLO_CO_UPsi2}.
The first complete calculation of NLO QCD corrections to the yield and polarization of $\Upsilon(1S,2S,3S)$ was presented in Ref.~\cite{NLO_CO_CS_YnS_First}, where the predicted yields were found to agree well with CDF~\cite{CDF_Upsi_Pol_1p8,CDF_Upsi_Pol} and CMS~\cite{CMS_Upsi_Pol} measurements over a wide $p_{\rm T}$ range. For $\Upsilon(1S,2S)$, the predicted polarizations are consistent with CMS data but show discrepancies with the CDF results. However, the polarization of $\Upsilon(3S)$ cannot be explained at LO in $v^2$ and NLO in $\alpha_s$ if feed-down contributions from higher excited bottomonium states, $\chi_{bJ}(3P)$,  are neglected. Subsequently, the mass of $\chi_{bJ}(3P)$ was measured by LHCb Collaboration~\cite{LHCb_Mass_chi3P}, and the fraction of $\Upsilon(3S)$ originating from $\chi_{bJ}(3P)$ radiative decay was first determined by the LHCb Collaboration~\cite{LHCb_Mass_chi3P_Y3S}. The sizable value of this contribution indicated the need to revisit the production and polarization of $\Upsilon(3S)$, as well as $\Upsilon(1S,2S)$. Following these developments, two studies~\cite{NLO_UPsi_chi_b_3P,NLO_UPsi_chi_b_3P2} incorporating the $\chi_{bJ}(3P)$ feed-down achieved a satisfactory description of the $\Upsilon$ polarization results at LHC. In Ref~\cite{NLO_UPsi_All}, a comprehensive analysis of $\Upsilon(1S,2S,3S)$ polarization is performed. All three polarization parameters, $\lambda_{\theta}$, $\lambda_{\theta\phi}$, and $\lambda_{\phi}$, for $\Upsilon(1S,2S,3S)$ hadroproduction are evaluated at QCD NLO within the NRQCD framework in both the helicity and Collins-Soper frames. For $\lambda_{\theta}$ and $\lambda_{\theta\phi}$, theoretical results provide a good description of the CMS measurements~\cite{CMS_Upsi_Pol} in both the helicity and Collins--Soper frames. In the case of LHCb data~\cite{LHCb_Upsi_Pol}, although most points are reasonably well reproduced, some discrepancies remain within the $3\sigma$ level. The agreement deteriorates for $\lambda_{\phi}$.



Despite significant progress, key theoretical challenges persist. Different NLO global fits yield inconsistent CO LDMEs, questioning their universality, while a simultaneous description of yields and polarization remains unresolved. Additional uncertainties from scale choices, feed-down contributions, and higher-order effects continue to complicate precise interpretations, highlighting the need for transparent baseline calculations.


In this work, we employ LO NRQCD as it provides a transparent and well-controlled framework for isolating individual production channels and feed-down contributions. The LO calculations allow us to isolate the contributions of individual CS and CO mechanisms and to systematically study their dependence on the choice of
LDMEs. Importantly, both LO and NLO calculations are known to describe the measured quarkonium production cross-sections reasonably well. While NLO corrections play a crucial role in resolving polarization observables where LO predictions are inadequate. The present study is focused solely on cross-sections, for which LO already captures the essential physics. This makes LO NRQCD a suitable and efficient choice, offering analytical clarity and a robust baseline for future extensions to more complex environments such as proton–nucleus and heavy-ion collisions.

Experimentally, bottomonium production in $pp$ collisions has been extensively measured over a wide range of center-of-mass energies, $\sqrt{s}$ = 2.76, 5.02, 7, 8, and 13 TeV, by the ALICE~\cite{ALICE_Upsilon_pp_7TeV,ALICE_Upsilon_pp_8TeV}, ATLAS~\cite{ATLAS_Upsilon_pp_7TeV,ATLAS_Upsilon_pp_7TeV2}, CMS~\cite{CMS_Upsilon_pp_7TeV,CMS_Upsilon_pp_8TeV,CMS_Upsilon_pp_13TeV}, and LHCb~\cite{LHCb_Upsilon_pp_2.76TeV,LHCb_Upsilon_pp_5TeV,LHCb_Upsilon_pp_7TeV,LHCb_Upsilon_pp_8TeV,LHCb_Upsilon_pp_13TeV}  Collaborations at the LHC, covering forward, near-forward, and midrapidity regions. These measurements consistently report prompt $\Upsilon(nS)$ production cross-sections, which include both direct production and feed-down contributions from higher excited bottomonium states. Motivated by this, we compute the $\Upsilon(nS)$ production cross-sections and cross-section ratios among different $\Upsilon$ states at $\sqrt{s} = 7$ and $13$ TeV within the LO NRQCD framework and perform a detailed comparison with available LHC data. 




The paper is organized as follows. Section 2 outlines the theoretical framework based on NRQCD. Section 3 presents the results and their comparison with experimental data, followed by the summary and discussion in Section 4.

\section{Theoretical formalism}
We study prompt $\Upsilon(nS)$ production in $pp$ collisions within the NRQCD factorization framework. The production of a physical bottomonium state is expressed as a sum over the perturbative production of a $b\bar b$ pair in a definite Fock state and its subsequent nonperturbative transition into the quarkonium state,


\begin{eqnarray}
&d\sigma_{A+B\rightarrow H+X} = \sum_{a,b,n}\int dx_a\,dx_b\,
G_{a/A}(x_a,\mu_F^2)G_{b/B}(x_b,\mu_F^2)\nonumber
\\& \times d\hat{\sigma}
(ab\to b\bar b[n]+X)
\langle {\cal O}^{H}(n)\rangle
\end{eqnarray}


Here, $G_{a/A}$ and $G_{b/B}$ are the parton distribution functions
(PDFs), $\mu_F$ is the factorization scale, and
$n={}^{2S+1}L_J^{[c]}$ specifies the quantum numbers and color
representation of the intermediate $b\bar b$ pair. The short-distance
coefficients (SDCs) are calculable in perturbative QCD, whereas
$\langle {\cal O}^{H}(n)\rangle$ are the nonperturbative LDMEs. For a state of mass $m_H$ and transverse
momentum $p_T$, we define
\begin{equation}
m_T=\sqrt{p_T^2+m_H^2}.
\label{eq:mt}
\end{equation}

At leading order (LO), the relevant $2\to2$ partonic processes give
the differential cross section
%

\begin{eqnarray}
&\frac{{d^{2}\sigma}^{ab\rightarrow cd}}{dp_T\,dy} = \int dx_a\, G_{a/A}(x_a,\,\mu^{2}_{F})\, G_{b/B}(x_b,\,\mu^{2}_{F})\nonumber \\&\times 2p_T \frac{x_a\,x_b}{x_a-\frac{m_T}{\sqrt{s}}e^y}\frac{d \hat \sigma}{d\hat t},
\end{eqnarray}
where $\sqrt{s}$ is the total energy in the centre-of-mass and $y$ is the rapidity of the $b\bar b$ pair. The momentum fraction $x_b$ is fixed by kinematics,
\begin{eqnarray}
x_b = \frac{1}{\sqrt{s}}\frac{x_a\,\sqrt{s}\,m_T\,e^{-y}-m^2_H}{x_a\,\sqrt{s}-m_T\,e^y}.
\end{eqnarray}
with
\begin{eqnarray}
x_{a,\rm min} = \frac{1}{\sqrt{s}}\frac{\sqrt{s}\,m_T\,e^{y}-m^2_H}{\sqrt{s}-m_T\,e^{-y}}.
\end{eqnarray}
The partonic cross section is obtained from
\begin{equation}
\frac{d\hat{\sigma}}{d\hat t}
=
\frac{|\mathcal{M}|^2}{16\pi\hat{s}^{\,2}},
\label{eq:partonic_xsection}
\end{equation}
where $\hat{s}$ and $\hat{t}$ denote the partonic Mandelstam
variables. $\mathcal{M}$ is the Feynman amplitude for the process which contains $\alpha_{s}$. The strong coupling is evaluated at the renormalization scale $\mu_R$ using~\cite{npb469}
\begin{equation}
\alpha_s(\mu_R^2)
=
\frac{12\pi}
{(33-2N_f)\ln(\mu_R^2/\Lambda_{\rm QCD}^2)} .
\label{eq:alphas}
\end{equation}

For direct $\Upsilon(nS)$ production, the relevant NRQCD channels
are the color-singlet ${}^{3}S_1^{[1]}$ and the color-octet
${}^{1}S_0^{[8]}$, ${}^{3}S_1^{[8]}$, and ${}^{3}P_J^{[8]}$
states. Thus,
\begin{equation}
\begin{aligned}
d\sigma_{\rm dir}^{\Upsilon(nS)}
={}&
d\hat{\sigma}({}^{3}S_1^{[1]})
\langle{\cal O}^{\Upsilon(nS)}
({}^{3}S_1^{[1]})\rangle
\\
&+d\hat{\sigma}({}^{1}S_0^{[8]})
\langle{\cal O}^{\Upsilon(nS)}
({}^{1}S_0^{[8]})\rangle
\\
&+d\hat{\sigma}({}^{3}S_1^{[8]})
\langle{\cal O}^{\Upsilon(nS)}
({}^{3}S_1^{[8]})\rangle
\\
&+d\hat{\sigma}({}^{3}P_J^{[8]})
\langle{\cal O}^{\Upsilon(nS)}
({}^{3}P_J^{[8]})\rangle .
\end{aligned}
\label{eq:direct_upsilon}
\end{equation}
The SDCs for the color-singlet and color-octet channels are taken
from Refs.~\cite{zpc19,npb291,plb184} and
Refs.~\cite{prd53,prd53a}, respectively. The present calculation is
performed at LO in $\alpha_s$.

The direct production of $\chi_{bJ}(nP)$ involves the
${}^{3}P_J^{[1]}$ color-singlet and ${}^{3}S_1^{[8]}$ color-octet
channels,
\begin{equation}
\begin{aligned}
d\sigma_{\rm dir}^{\chi_{bJ}(nP)}
={}&
d\hat{\sigma}({}^{3}P_J^{[1]})
\langle{\cal O}^{\chi_{bJ}(nP)}
({}^{3}P_J^{[1]})\rangle
\\
&+
d\hat{\sigma}({}^{3}S_1^{[8]})
\langle{\cal O}^{\chi_{bJ}(nP)}
({}^{3}S_1^{[8]})\rangle .
\end{aligned}
\label{eq:direct_chib}
\end{equation}
Heavy-quark spin symmetry is used to relate the members of a given
P-wave multiplet,
\begin{equation}
\langle{\cal O}^{\chi_{bJ}(nP)}
({}^{3}S_1^{[8]})\rangle
=
(2J+1)
\langle{\cal O}^{\chi_{b0}(nP)}
({}^{3}S_1^{[8]})\rangle .
\label{eq:HQSS}
\end{equation}


The experimentally measured prompt yield contains both direct
production and feed-down from higher bottomonium states. We therefore
write
\begin{equation}
d\sigma_{\rm prompt}^{H}
=
d\sigma_{\rm dir}^{H}
+
\sum_{H'}
d\sigma_{\rm dir}^{H'}
{\cal B}(H'\to H+X),
\label{eq:prompt}
\end{equation}
%

For $\Upsilon(1S)$, feed-down from
$\Upsilon(2S)$, $\Upsilon(3S)$, $\chi_{bJ}(1P)$, and
$\chi_{bJ}(2P)$ is included. For $\Upsilon(2S)$, contributions from
$\Upsilon(3S)$ and $\chi_{bJ}(2P)$ are considered. In addition,
$\chi_{bJ}(3P)$ feed-down to $\Upsilon(3S)$ is explicitly included.
The masses and branching fractions used in the calculation are taken
from the latest Particle Data Group publication~\cite{PDG2026} and are summarized in
Table~\ref{BR_Upsilon}. The feed-down contributions from the $\chi_{bJ}(3P)$ states are also
considered for $\Upsilon(3S)$ production. In the present calculation,
we include the contributions from the experimentally identified
$\chi_{b1}(3P)$ and $\chi_{b2}(3P)$ states, while the
$\chi_{b0}(3P)$ state is not included because its mass has not yet
been experimentally measured. The corresponding
$\chi_{b1,2}(3P)\rightarrow\Upsilon(3S)+\gamma$ branching fractions
are taken from the theoretical calculation of
Ref.~\cite{Han2016}. This common feed-down prescription is applied consistently to all
three $\Upsilon(nS)$ states. It is important for the present analysis
because feed-down modifies not only the absolute normalization but
also the relative $p_T$ dependence of the three states. The production
ratios discussed below are therefore constructed from the complete
prompt yields.

The color-singlet LDMEs of the S-wave states are obtained from the
corresponding radial wave functions following
Refs.~\cite{CS_LDMEs,prc87}. The P-wave color-singlet matrix elements
are obtained analogously. The numerical values are listed in
Table~\ref{LDMEs_Upsilon}. 

The color-octet LDMEs are not fixed by the leading color-singlet wave
functions and are determined phenomenologically. We extract them by
comparing the calculated $p_T$-differential cross sections with
measurements from CDF~\cite{CDF_Upsi_Pol_1p8}, ALICE~\cite{ALICE_Upsilon_pp_7TeV,ALICE_Upsilon_pp_8TeV}, ATLAS~\cite{ATLAS_Upsilon_pp_7TeV,ATLAS_Upsilon_pp_7TeV2}, CMS~\cite{CMS_Upsilon_pp_7TeV,CMS_Upsilon_pp_8TeV,CMS_Upsilon_pp_13TeV}, and LHCb~\cite{LHCb_Upsilon_pp_7TeV,LHCb_Upsilon_pp_8TeV,LHCb_Upsilon_pp_13TeV} in $pp$ collisions at $\sqrt{s}$ = 1.8, 7, 8 and 13 TeV. Separate fits are performed sequentially for
$\Upsilon(3S)$, $\Upsilon(2S)$, and $\Upsilon(1S)$ production.
Correlations among the fitted parameters are included through the
covariance-matrix method of Ref.~\cite{NLO_JPsi_Pol3}.

Some of the fitted color-octet LDMEs listed in
Table~\ref{LDMEs_Upsilon} have negative central values. Such values
do not represent negative probabilities. In the NRQCD factorization
formula, individual LDMEs multiply channel-dependent short-distance
coefficients and are not themselves observable production
probabilities. Their fitted values can therefore have either sign,
particularly when different channels are correlated. The physical
cross section is obtained only after summing all contributing
channels. The negative values encountered here are consequently
interpreted as phenomenological fit parameters rather than as
probabilities of individual production mechanisms.


We consider four main sources of theoretical uncertainty in the
calculated $\Upsilon(nS)$ production cross sections: the variation of
the renormalization and factorization scales, the bottom-quark mass,
the choice of PDFs, and the uncertainties in the branching fractions
of the feed-down decays. 

The perturbative uncertainty is estimated by varying the
renormalization and factorization scales around the characteristic
scale
\begin{equation}
\mu_0=\sqrt{p_T^2+m_b^2},
\end{equation}
according to
\begin{equation}
\mu_R=\xi_R\mu_0,\qquad
\mu_F=\xi_F\mu_0,
\end{equation}
where the central prediction corresponds to $\xi_R=\xi_F=1$.
The two scales are varied independently within
$0.5\leq\xi_R,\xi_F\leq2$, with the additional restriction
$0.5\leq\xi_R/\xi_F\leq2$. 
The scale variation gives
the dominant uncertainty, reaching approximately $42\%$ in the
lowest transverse-momentum interval,
$4<p_T<5$~GeV.

The dependence on the bottom-quark mass is studied by varying
$m_b$ around the central value 
$m_b$ = 4.75~{\rm GeV} 
within the range $4.5\leq m_b\leq5.0$~GeV. The resulting variation
of the cross section is found to be about $11\%$ at most, with the
largest effect occurring at low $p_T$.

The uncertainty associated with the proton structure is evaluated by
repeating the calculation with modern leading-order PDF
parameterizations. The central calculation employs the
NNPDF4.0 LO set~\cite{NNPDF40},
while the spread obtained from the alternative LO PDF sets is used
to estimate the PDF uncertainty. The PDF uncertainty is particularly
relevant at low and intermediate transverse momenta.


The uncertainties associated with the branching fractions entering
the feed-down contributions are also propagated to the prompt
$\Upsilon(nS)$ cross sections. Among the decay channels considered,
the uncertainty associated with
$\chi_{b0}(2P)\rightarrow\Upsilon(1S)+\gamma$ is one of the largest,
amounting to approximately $45\%$. Since the scale, mass, PDF, and branching-fraction uncertainties originate from different inputs, they are treated as independent and combined according to
\begin{equation}
\Delta_{\rm tot}
=
\sqrt{
\Delta_{\mu}^{\,2}
+\Delta_{m_b}^{\,2}
+\Delta_{\rm PDF}^{\,2}
+\Delta_{\rm BR}^{\,2}
}.
\label{eq:total_uncertainty}
\end{equation}


For the ratios of prompt $\Upsilon$ cross sections, a significant part of the common theoretical uncertainty cancels between the numerator
and denominator. This cancellation is expected for quantities calculated with the same
PDFs, heavy-quark mass, renormalization and factorization scales, and short-distance
production framework. Consequently, the ratios are less sensitive to overall normalization
uncertainties.

\begin{table*}
\centering
\renewcommand{\arraystretch}{1.2}
\setlength{\tabcolsep}{4.5pt}
\begin{tabular}{lccccccccccc}
\hline\hline
State $H$ &
$\Upsilon(1S)$ & $\Upsilon(2S)$ & $\Upsilon(3S)$ &
$\chi_{b0}(1P)$ & $\chi_{b1}(1P)$ & $\chi_{b2}(1P)$ &
$\chi_{b0}(2P)$ & $\chi_{b1}(2P)$ & $\chi_{b2}(2P)$ & $\chi_{b1}(3P)$ & $\chi_{b2}(3P)$\\
\hline
$\mathcal{B}(H\!\to\!\mu^{+}\mu^{-})$ [\%] &
2.48 & 1.93 & 2.18 &
-- & -- & -- & -- & -- & -- & -- & --\\
$\mathcal{B}(H\!\to\!\Upsilon(1S))$ [\%] &
-- & 26.5 & 6.6 &
1.94 & 35.2 & 18.0 &
0.4 & 9.9 & 6.6 & -- & --\\
$\mathcal{B}(H\!\to\!\Upsilon(2S))$ [\%] &
-- & -- & 10.6 &
-- & -- & -- &
1.38 & 18.1 & 8.9 & -- & --\\
$\mathcal{B}(H\!\to\!\Upsilon(3S))$ [\%] &
-- & -- & -- &
-- & -- & -- &
-- & -- & -- & 10.44 & 6.11 \\
$M_{H}$ [GeV] &
9.460 & 10.023 & 10.355 &
9.859 & 9.893 & 9.912 &
10.233 & 10.255 & 10.269 & 10.513 & 10.524\\
\hline\hline
\end{tabular}
\caption{\label{BR_Upsilon}
Branching ratios and masses of bottomonia are taken from PDG~\cite{PDG2026} and~\cite{Han2016}.}
\end{table*}

 \begin{table*}
\centering
\begin{tabular}{l|cccc}
\hline
\hline
{~}&{~~~~~~~~~~~~~~~~~~~~~~~}&{~~~~~~~~~~~~~~~~~~~~}&{NRQCD}\\
&LDMEs&Numerical&scaling\\
&~~~&value&order\\
\hline
&~$<\mathcal{O}(Q\bar Q(~^3S_1^{[1]})\rightarrow \Upsilon(1S))>$ & 10.9 GeV$^{3}$& $m_{b}^{3}v_{b}^{3}$ \\  
&~$<\mathcal{O}(Q\bar Q(~^3S_1^{[1]})\rightarrow \Upsilon(2S))>$ & 4.5 GeV$^{3}$& $m_{b}^{3}v_{b}^{3}$ \\ 
Colour-&~$<\mathcal{O}(Q\bar Q(~^3S_1^{[1]})\rightarrow \Upsilon(3S))>$ & 4.3 GeV$^{3}$& $m_{b}^{3}v_{b}^{3}$ \\ 
Singlet&~$<\mathcal{O}(Q\bar Q(~^3P_{J}^{[1]}) \rightarrow \chi_{bJ}(1P))>$/$(2J+1)$$m_{b}^{2}$ & 0.100 GeV$^{3}$& $m_{b}^{3}v_{b}^{5}$ \\ 
&~$<\mathcal{O}(Q\bar Q(~^3P_{J}^{[1]}) \rightarrow \chi_{bJ}(2P))>$/$(2J+1)$$m_{b}^{2}$ & 0.036 GeV$^{3}$& $m_{b}^{3}v_{b}^{5}$ \\

&~$<\mathcal{O}(Q\bar Q(~^3P_{J}^{[1]}) \rightarrow \chi_{bJ}(3P))>$/$(2J+1)$$m_{b}^{2}$ & 0.122 GeV$^{3}$& $m_{b}^{3}v_{b}^{5}$ \\
\hline
&~$<\mathcal{O}(Q\bar Q(~^3S_1^{[8]})\rightarrow \Upsilon(1S))>$ & - 0.0042 $\pm$ 0.0025 GeV$^{3}$& $m_{b}^{3}v_{b}^{7}$ \\
&~$<\mathcal{O}(Q\bar Q(~^3S_1^{[8]})\rightarrow \Upsilon(2S))>$ & 0.0032 $\pm$ 0.0075 GeV$^{3}$& $m_{b}^{3}v_{b}^{7}$ \\
&~$<\mathcal{O}(Q\bar Q(~^3S_1^{[8]})\rightarrow \Upsilon(3S))>$ & 0.0273 $\pm$ 0.0015 GeV$^{3}$& $m_{b}^{3}v_{b}^{7}$ \\
&~$<\mathcal{O}(Q\bar Q(~^3S_1^{[8]})\rightarrow \chi_{bJ}(1P))>$/$(2J+1)$$m_{b}^{2}$ & 0.0129 $\pm$ 0.0017 GeV$^{3}$& $m_{b}^{3}v_{b}^{5}$ \\
Colour-&~$<\mathcal{O}(Q\bar Q(~^3S_1^{[8]})\rightarrow \chi_{bJ}(2P))>$/$(2J+1)$$m_{b}^{2}$ & 0.0274 $\pm$ 0.0065 GeV$^{3}$& $m_{b}^{3}v_{b}^{5}$ \\

Octet&~$<\mathcal{O}(Q\bar Q(~^3S_1^{[8]})\rightarrow \chi_{bJ}(3P))>$/$(2J+1)$$m_{b}^{2}$ & 0.0300 $\pm$ 0.0070 GeV$^{3}$& $m_{b}^{3}v_{b}^{5}$ \\

&~$<\mathcal{O}(Q\bar Q(~^1S_0^{[8]})\rightarrow \Upsilon(1S))>$ & 0.1113 $\pm$ 0.0045 GeV$^{3}$& $m_{b}^{3}v_{b}^{7}$ \\
&~$<\mathcal{O}(Q\bar Q(~^1S_0^{[8]})\rightarrow \Upsilon(2S))>$ & 0.0358 $\pm$ 0.0214 GeV$^{3}$& $m_{b}^{3}v_{b}^{7}$ \\
&~$<\mathcal{O}(Q\bar Q(~^1S_0^{[8]})\rightarrow \Upsilon(3S))>$ & - 0.0109 $\pm$ 0.0106 GeV$^{3}$& $m_{b}^{3}v_{b}^{7}$ \\
&~$<\mathcal{O}(Q\bar Q(~^3P_0^{[8]})\rightarrow \Upsilon(1S))>$/$m_{b}^{2}$ & - 0.0063 $\pm$ 0.0001 GeV$^{3}$& $m_{b}^{3}v_{b}^{7}$ \\
&~$<\mathcal{O}(Q\bar Q(~^3P_0^{[8]})\rightarrow \Upsilon(2S))>$/$m_{b}^{2}$ & - 0.0058 $\pm$ 0.0050 GeV$^{3}$& $m_{b}^{3}v_{b}^{7}$ \\
&~$<\mathcal{O}(Q\bar Q(~^3P_0^{[8]})\rightarrow \Upsilon(3S))>$/$m_{b}^{2}$ & 0.0041 $\pm$ 0.0027 GeV$^{3}$& $m_{b}^{3}v_{b}^{7}$ \\
\hline
\hline
\end{tabular}
\caption{The colour-singlet and colour-octet matrix elements with numerical values and NRQCD scaling order for bottomonia.}
\label{LDMEs_Upsilon}
\end{table*}

\section{Results}
The differential production cross-sections of $\Upsilon(nS)$ in $pp$ collisions at $\sqrt{s}$ = 7 and 13 TeV have been evaluated within the framework of NRQCD factorization. The obtained results are compared with the corresponding experimental measurements reported by ATLAS ($|y|<1.2$ and $1.2<|y|<2.25$), CMS ($|y|<1.2$ and $|y|<2.4$), LHCb ($2<y<4.5$), and ALICE ($2.5<y<4$). This comprehensive comparison allows us to examine the validity of NRQCD predictions across mid-, near-forward-, and forward-rapidity regions at LHC energies.

In our NRQCD computations, the central predictions correspond to the renormalization and factorization scales $\xi_R=\xi_F=1$, while the upper and lower bounds of the theoretical uncertainty band are obtained by varying these scales to $\xi_R=\xi_F=0.5$ and $\xi_R=\xi_F=2$, respectively.

\begin{figure*}
    \centering
 \includegraphics[width=8.5cm,height=6.5cm,angle=0]{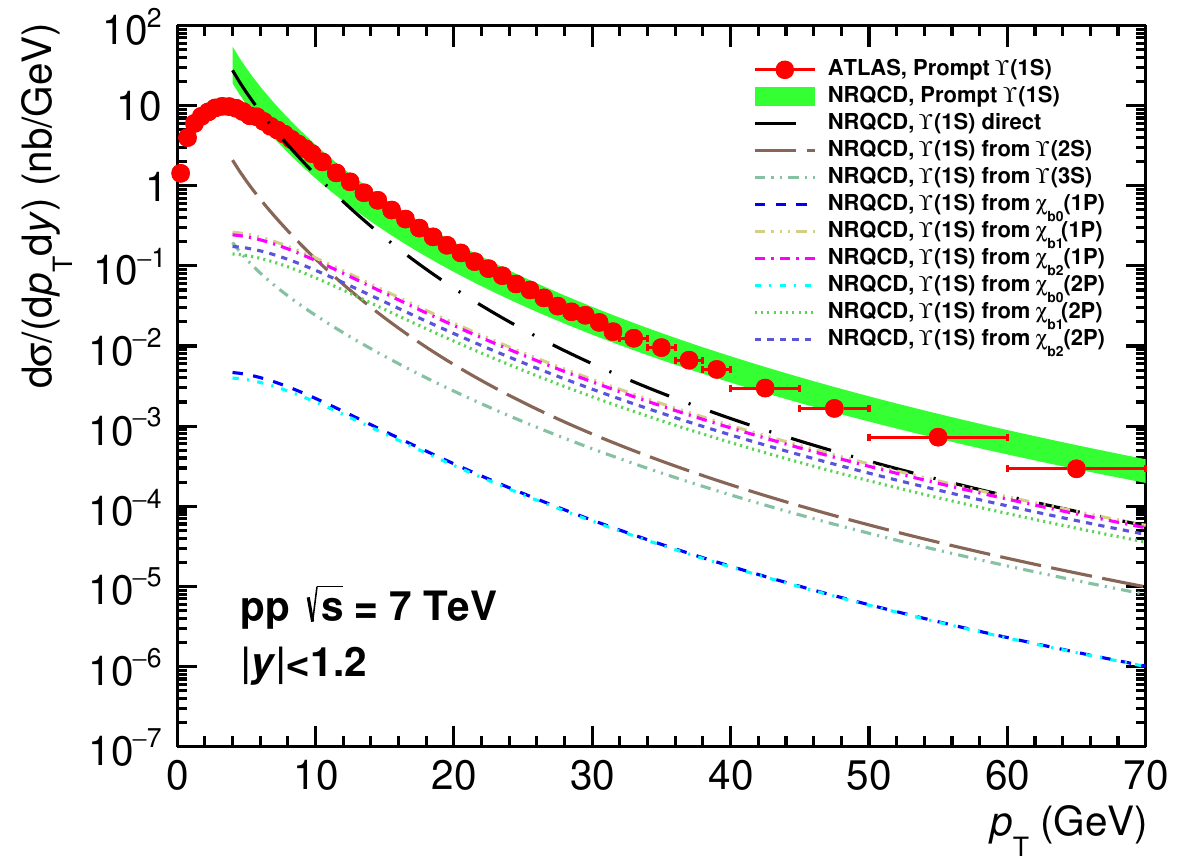}~~~~ \includegraphics[width=8.5cm,height=6.5cm,angle=0]{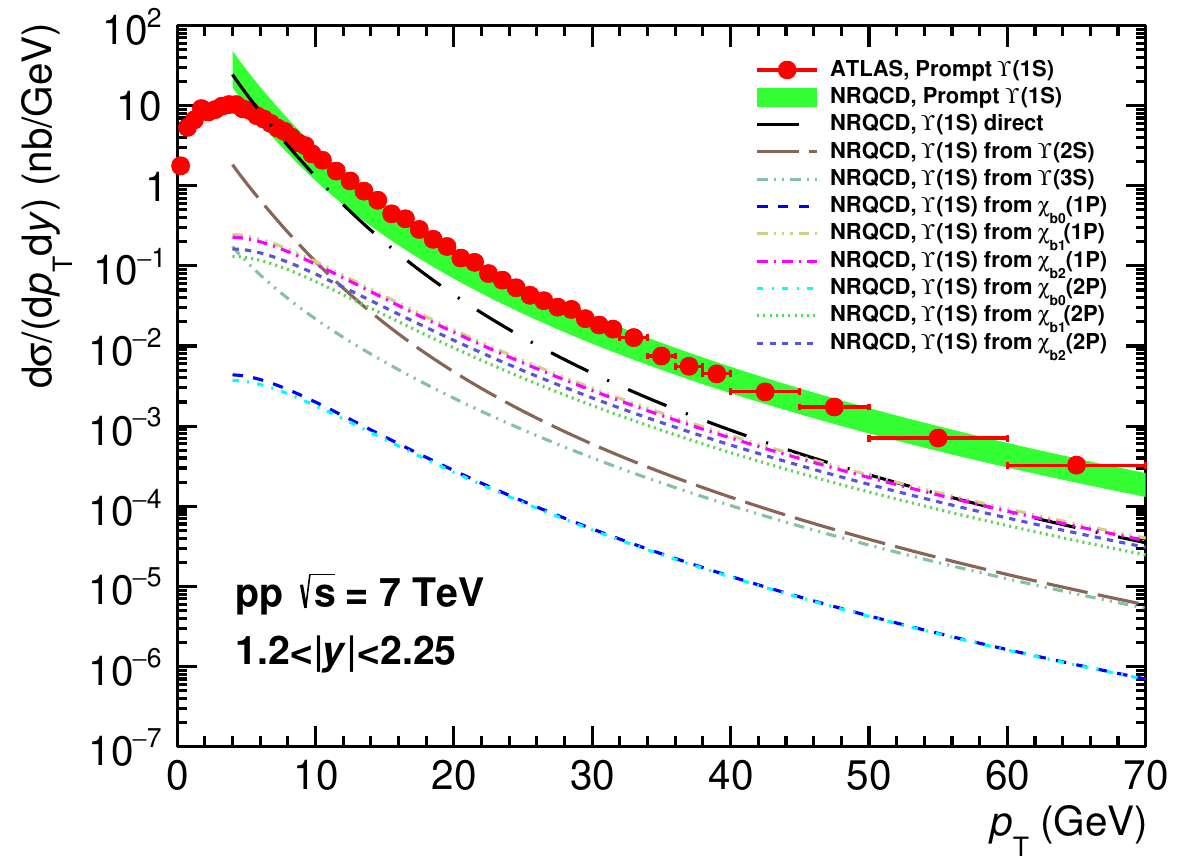}~~~~~\\  \includegraphics[width=8.5cm,height=6.5cm,angle=0]{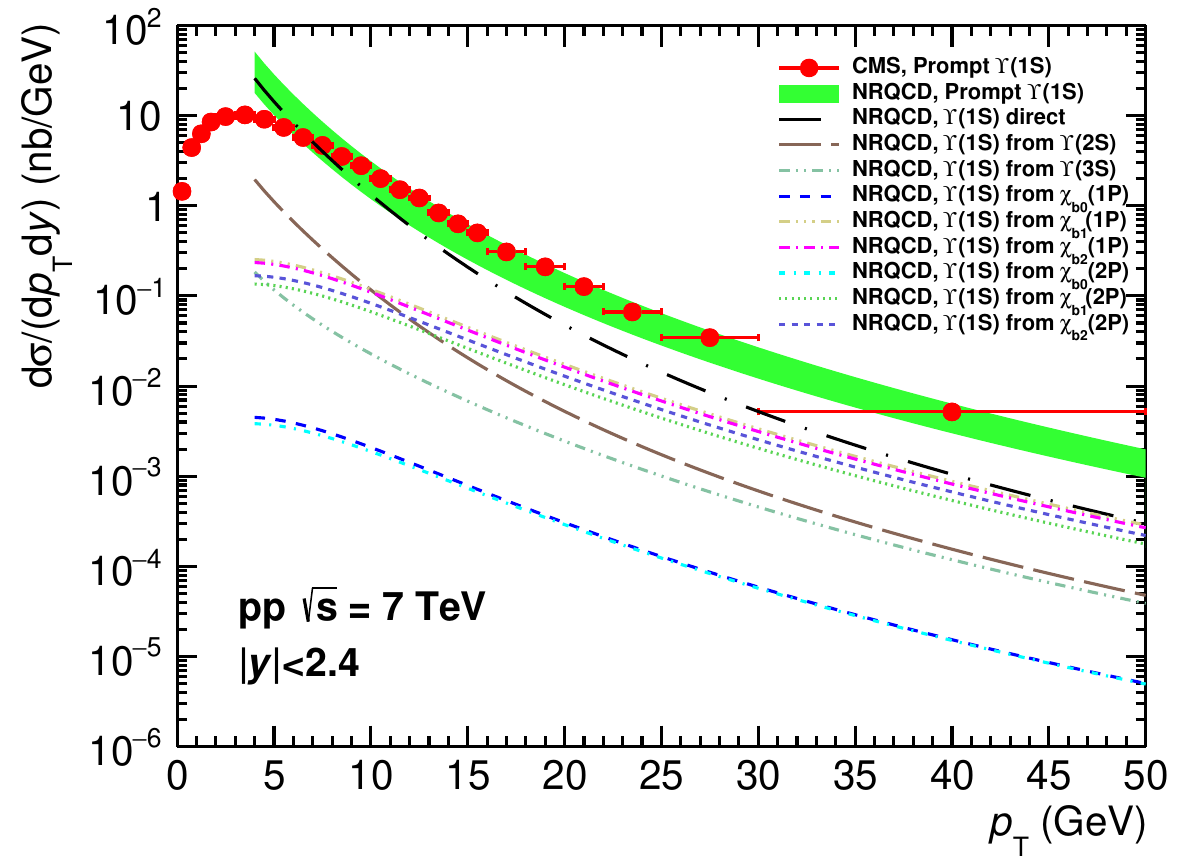}~~~~ \includegraphics[width=8.5cm,height=6.5cm,angle=0]{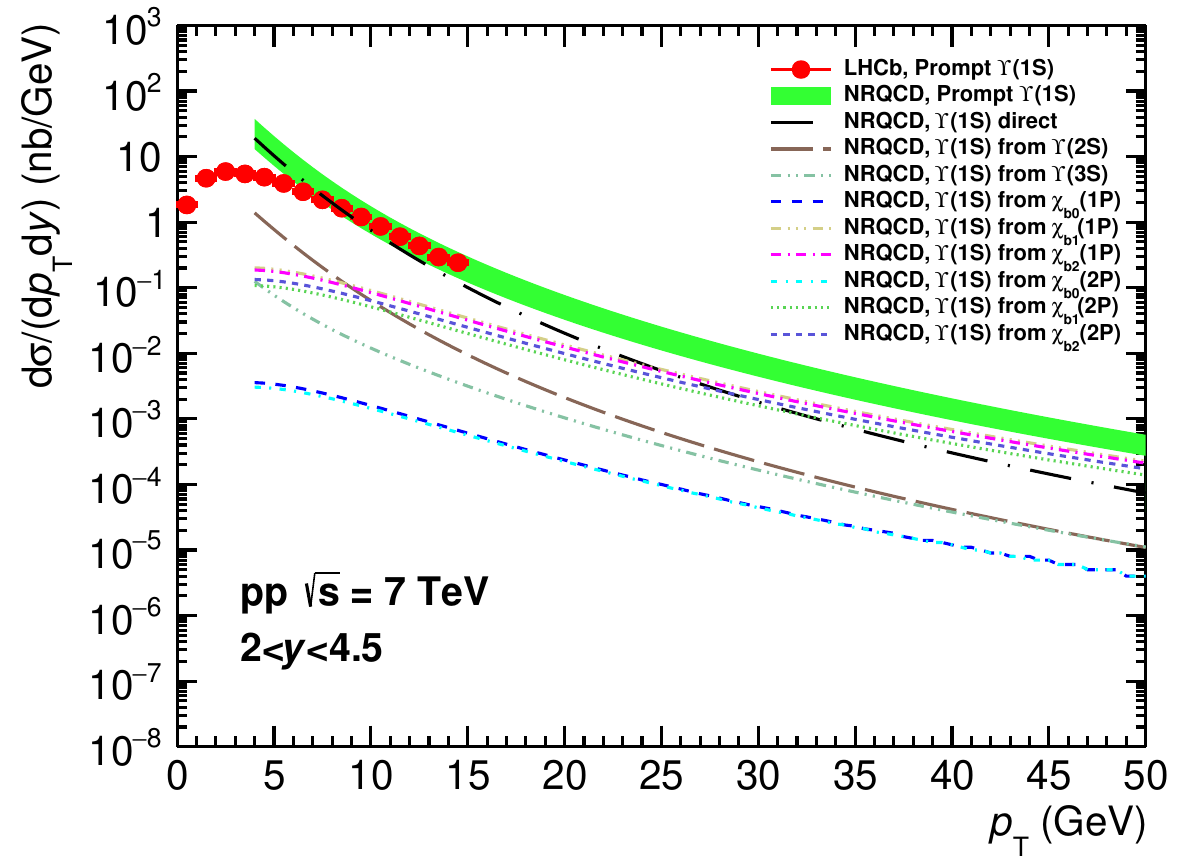}~~~~~\\ \includegraphics[width=8.5cm,height=6.5cm,angle=0]{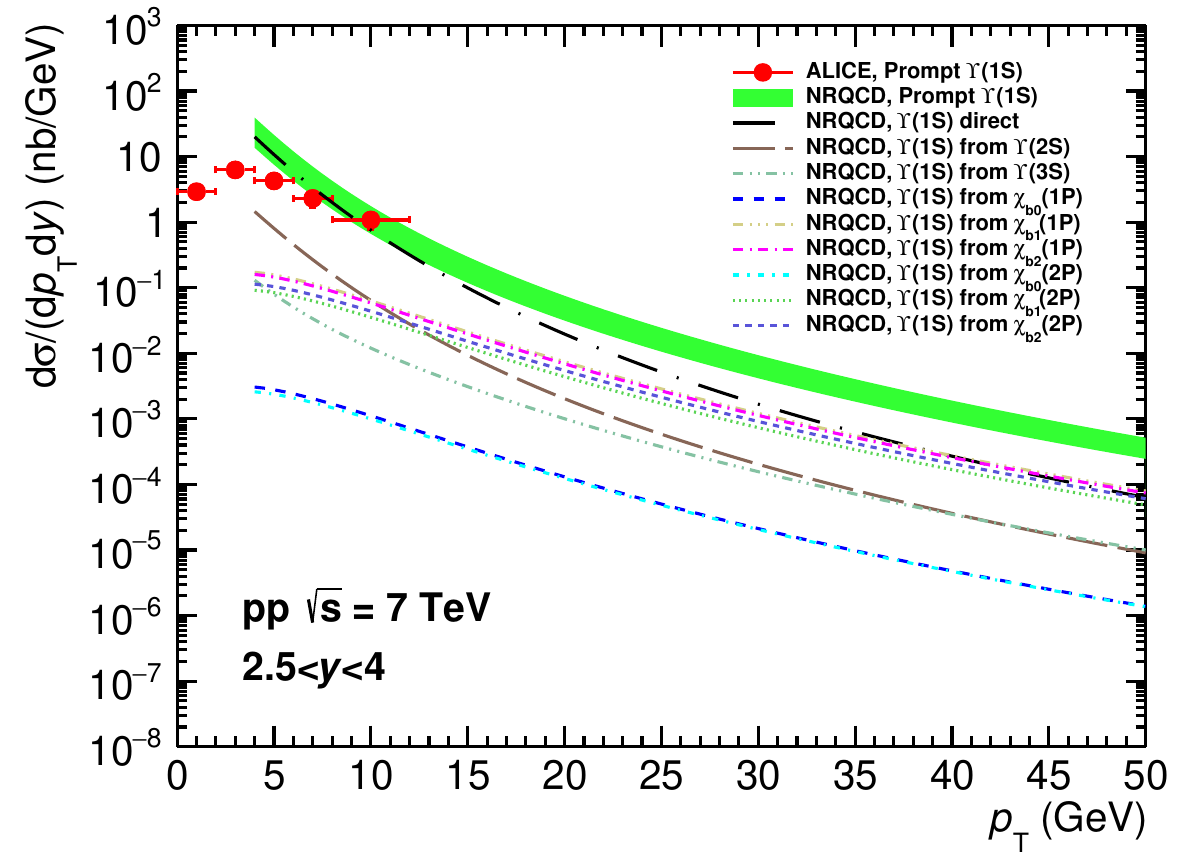}
 \caption{Differential production cross-section of $\Upsilon(1S)$ as a function of $p_{T}$ compared with the measurements by ATLAS~\cite{ATLAS_Upsilon_pp_7TeV}, CMS~\cite{CMS_Upsilon_pp_7TeV}, LHCb~\cite{LHCb_Upsilon_pp_7TeV} and ALICE~\cite{ALICE_Upsilon_pp_7TeV} in $pp$ collisions at $\sqrt{s}$ = 7 TeV. The vertical error bars are the quadratic sum of statistical and systematic uncertainties. The calculations corresponding to the sum of all contributions are shown as a green band. The direct and feed-down contributions to $\Upsilon(1S)$ are shown only by lines for the central values.}
 \label{Y1S_NRQCD}
\end{figure*}

Figure~\ref{Y1S_NRQCD} presents the differential cross-section of $\Upsilon(1S)$ as a function of $p_{\rm T}$, as obtained from our NRQCD calculations, compared with the experimental results from ATLAS~\cite{ATLAS_Upsilon_pp_7TeV}, CMS~\cite{CMS_Upsilon_pp_7TeV}, LHCb~\cite{LHCb_Upsilon_pp_7TeV}, and ALICE~\cite{ALICE_Upsilon_pp_7TeV} in $pp$ collisions at $\sqrt{s}=7$ TeV. In the experimental data, vertical error bars are the quadratic sum of statistical and systematic uncertainties. The ATLAS experiment measured the $\Upsilon(1S)$ production cross-section at mid-rapidity ($|y|<1.2$) and near-forward rapidity ($1.2<|y|<2.25$) up to $p_{\rm T}\approx70$ GeV, whereas CMS covered the mid-rapidity region ($|y|<2.4$) up to $p_{\rm T}\approx50$ GeV. The LHCb and ALICE measurements extend to forward rapidities ($2<y<4.5$ and $2.5<y<4$, respectively) with $p_{\rm T}$ reach up to about 15 GeV and 12 GeV. Our NRQCD results, representing the sum of all contributing processes, are displayed as a green uncertainty band. The individual central values corresponding to the direct $\Upsilon(1S)$ production and the feed-down contributions from $\Upsilon(2S)$, $\Upsilon(3S)$, $\chi_{bJ}(1P)$, and $\chi_{bJ}(2P)$ states are also shown separately as lines. The inclusion of these feed-down contributions is found to be essential for reproducing the correct $p_{\rm T}$-dependent shape of the prompt $\Upsilon(1S)$ cross-section. From Fig.~\ref{Y1S_NRQCD}, it can be observed that our NRQCD calculations provide good agreement with the experimental data from all four Collaborations, particularly for $p_{\rm T} > 4$ GeV. 

\begin{figure*}
    \centering
 \includegraphics[width=8.5cm,height=6.5cm,angle=0]{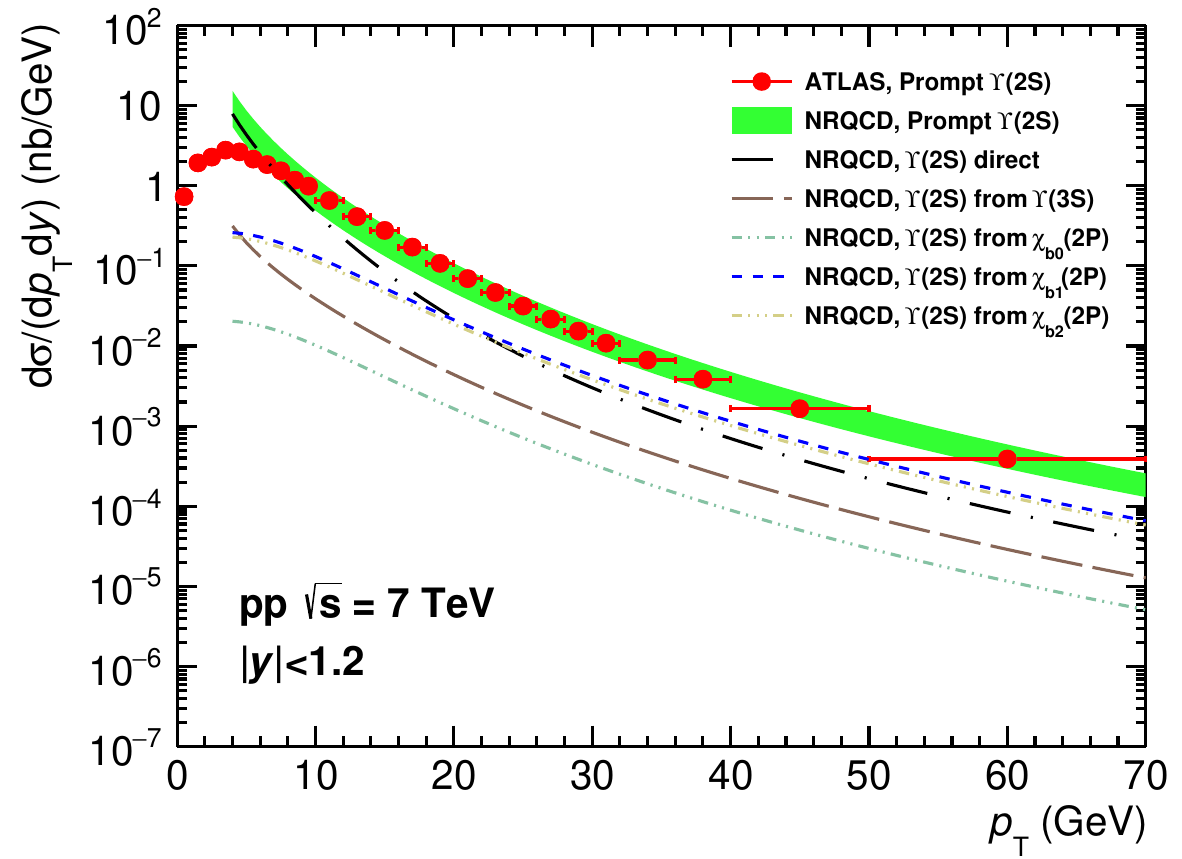}~~~~ \includegraphics[width=8.5cm,height=6.5cm,angle=0]{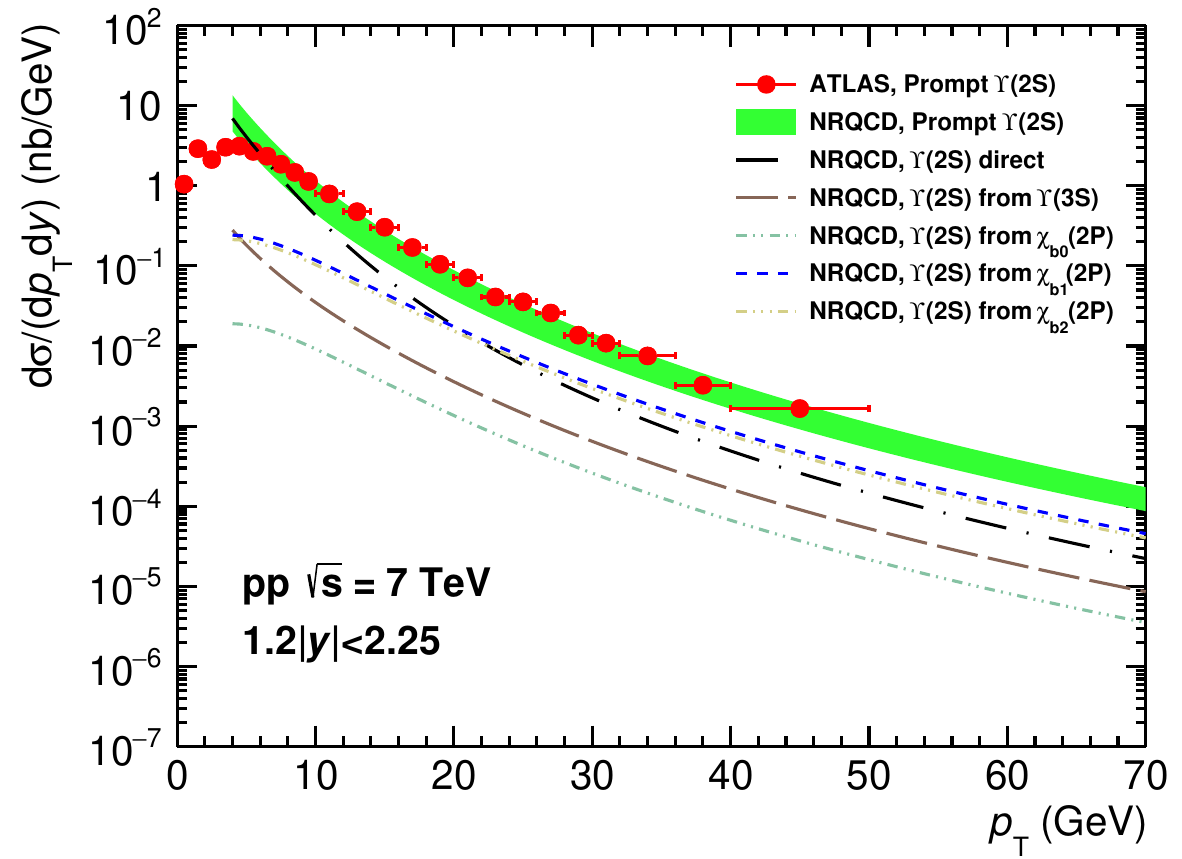}~~~~~\\  \includegraphics[width=8.5cm,height=6.5cm,angle=0]{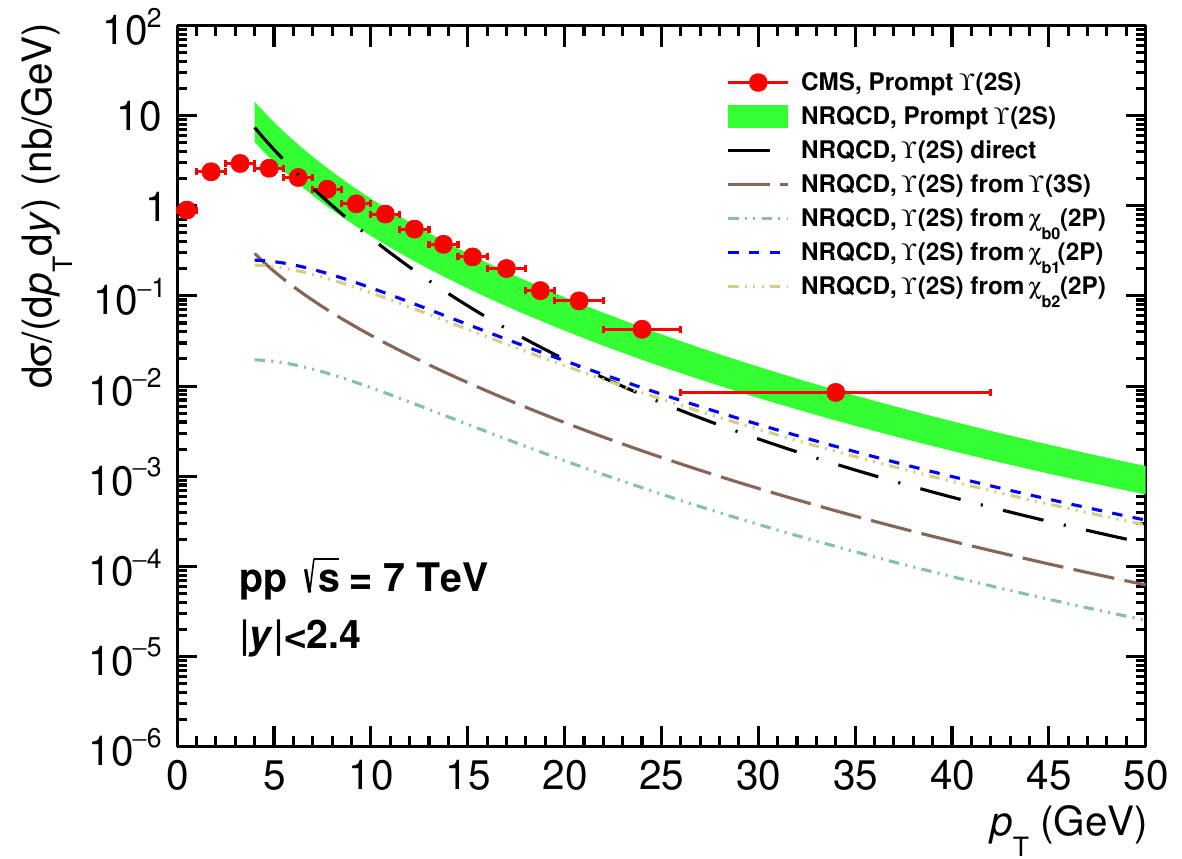}~~~~ \includegraphics[width=8.5cm,height=6.5cm,angle=0]{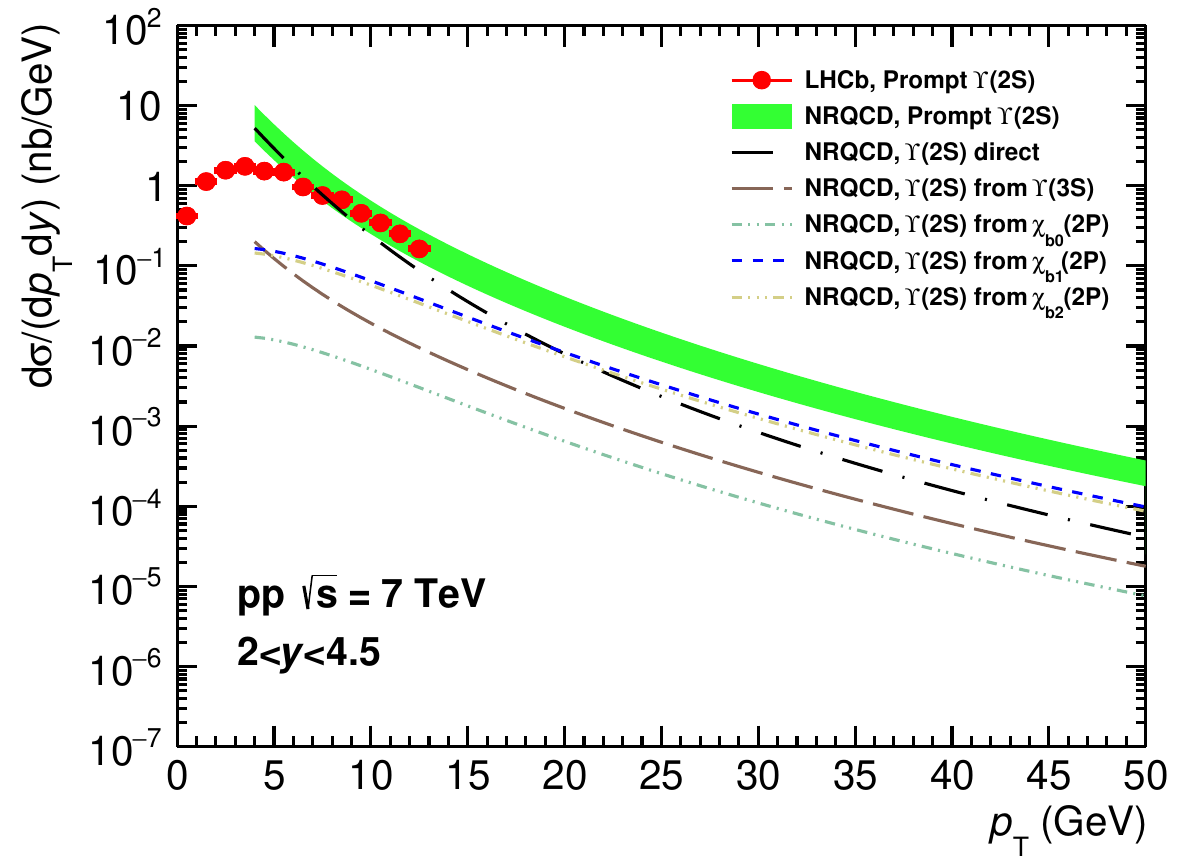}
 \caption{Differential production cross-section of $\Upsilon(2S)$ as a function of $p_{T}$ compared with the measurements by ATLAS~\cite{ATLAS_Upsilon_pp_7TeV}, CMS~\cite{CMS_Upsilon_pp_7TeV}, LHCb~\cite{LHCb_Upsilon_pp_7TeV} in p-p collisions at $\sqrt{s}$ = 7 TeV.}
 \label{Y2S_NRQCD}
 \end{figure*}

Figure~\ref{Y2S_NRQCD} shows the differential production cross-section of $\Upsilon(2S)$ as a function of $p_{\rm T}$ obtained from our NRQCD calculations, compared with the experimental data from ATLAS~\cite{ATLAS_Upsilon_pp_7TeV}, CMS~\cite{CMS_Upsilon_pp_7TeV}, and LHCb~\cite{LHCb_Upsilon_pp_7TeV} in $pp$ collisions at $\sqrt{s}=7$ TeV. The rapidity and $p_{\rm T}$ ranges for the $\Upsilon(2S)$ measurements in ATLAS and LHCb are identical to those as for $\Upsilon(1S)$. The CMS experiment measured $\Upsilon(2S)$ production in the mid-rapidity region ($|y|<2.4$) up to $p_{\rm T}\approx42$ GeV. In our NRQCD framework, the total $\Upsilon(2S)$ yield, including all relevant contributions, is shown as a green uncertainty band. The individual central values corresponding to direct $\Upsilon(2S)$ production and feed-down contributions from higher excited states, $\Upsilon(3S)$ and $\chi_{bJ}(2P)$, are displayed as separate lines. As seen from Fig.~\ref{Y2S_NRQCD}, the inclusion of these feed-down contributions is essential to accurately reproduce the $p_{\rm T}$-dependent shape of the prompt $\Upsilon(2S)$ cross-section. Furthermore, our NRQCD predictions exhibit good agreement with the ATLAS, CMS, and LHCb measurements, particularly for $p_{\rm T} > 4$ GeV.

\begin{figure*}
    \centering
 \includegraphics[width=8.5cm,height=6.5cm,angle=0]{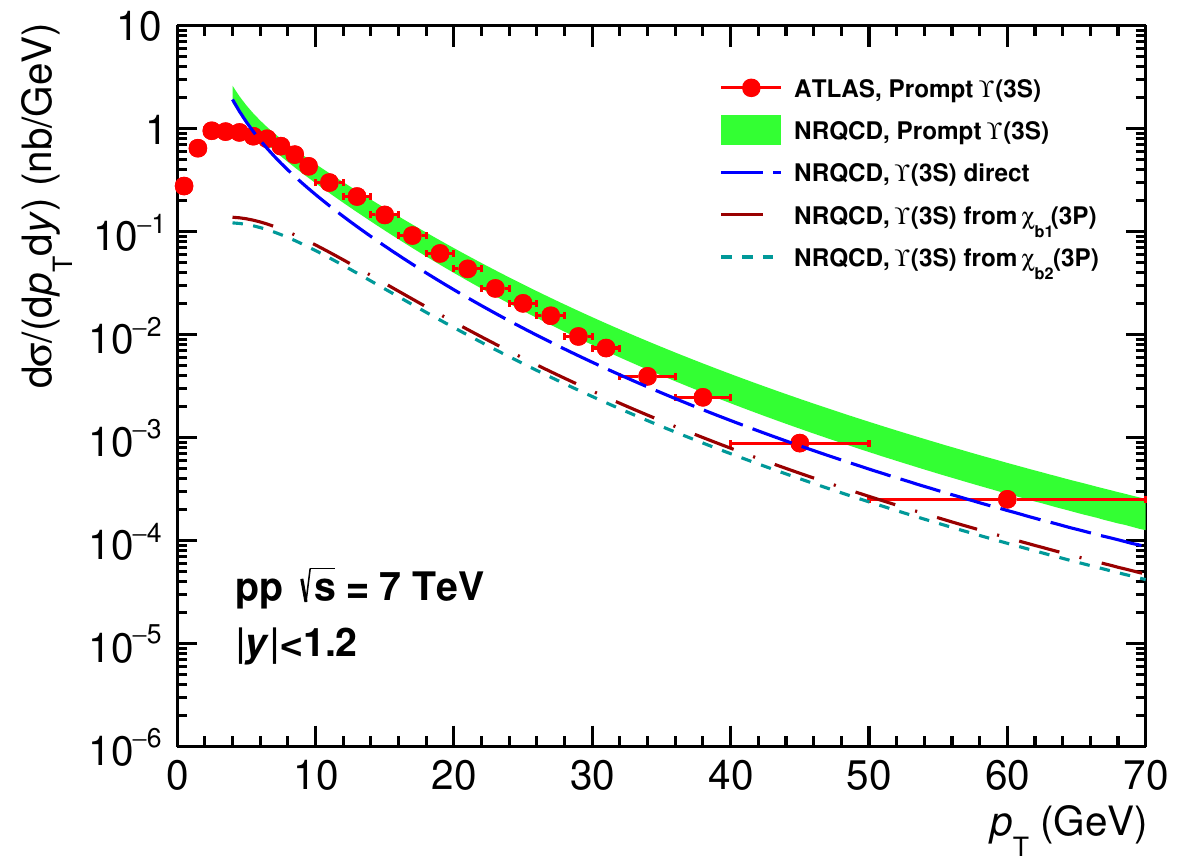}~~~~ \includegraphics[width=8.5cm,height=6.5cm,angle=0]{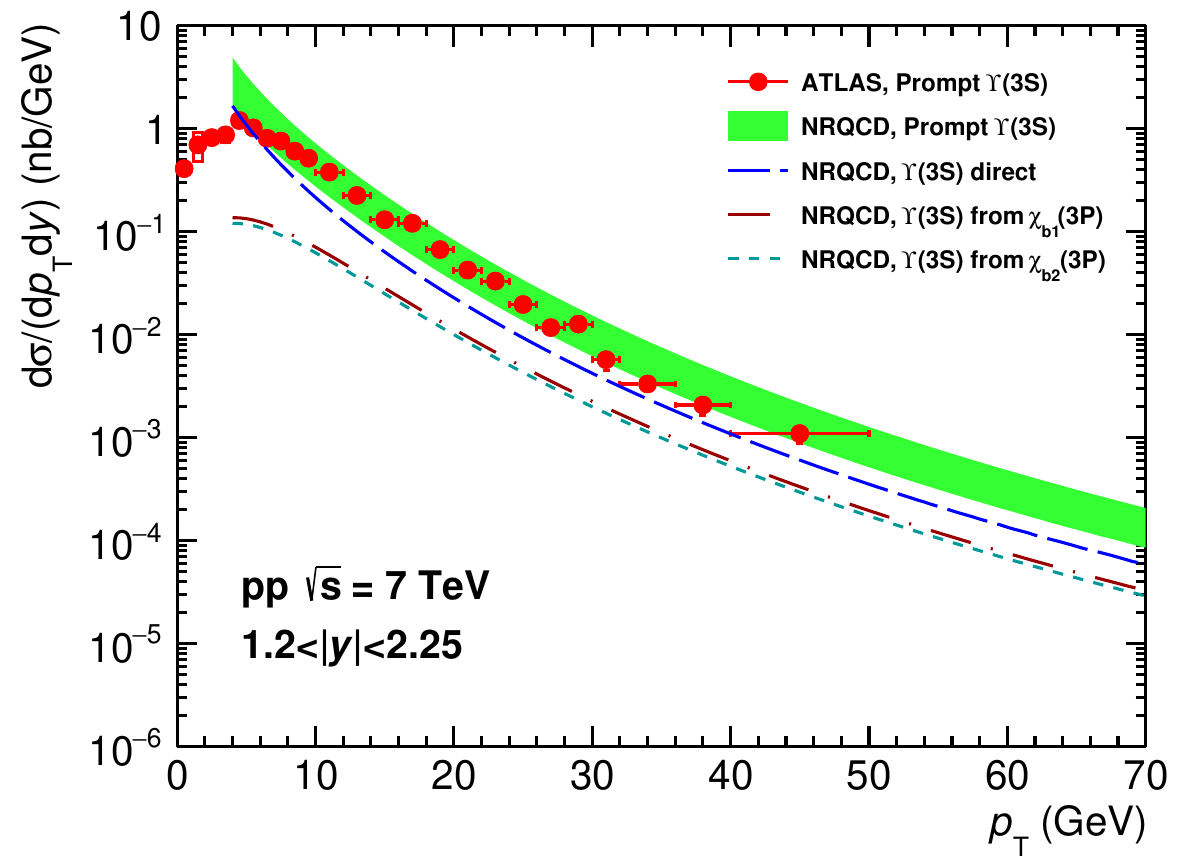}~~~~~\\  \includegraphics[width=8.5cm,height=6.5cm,angle=0]{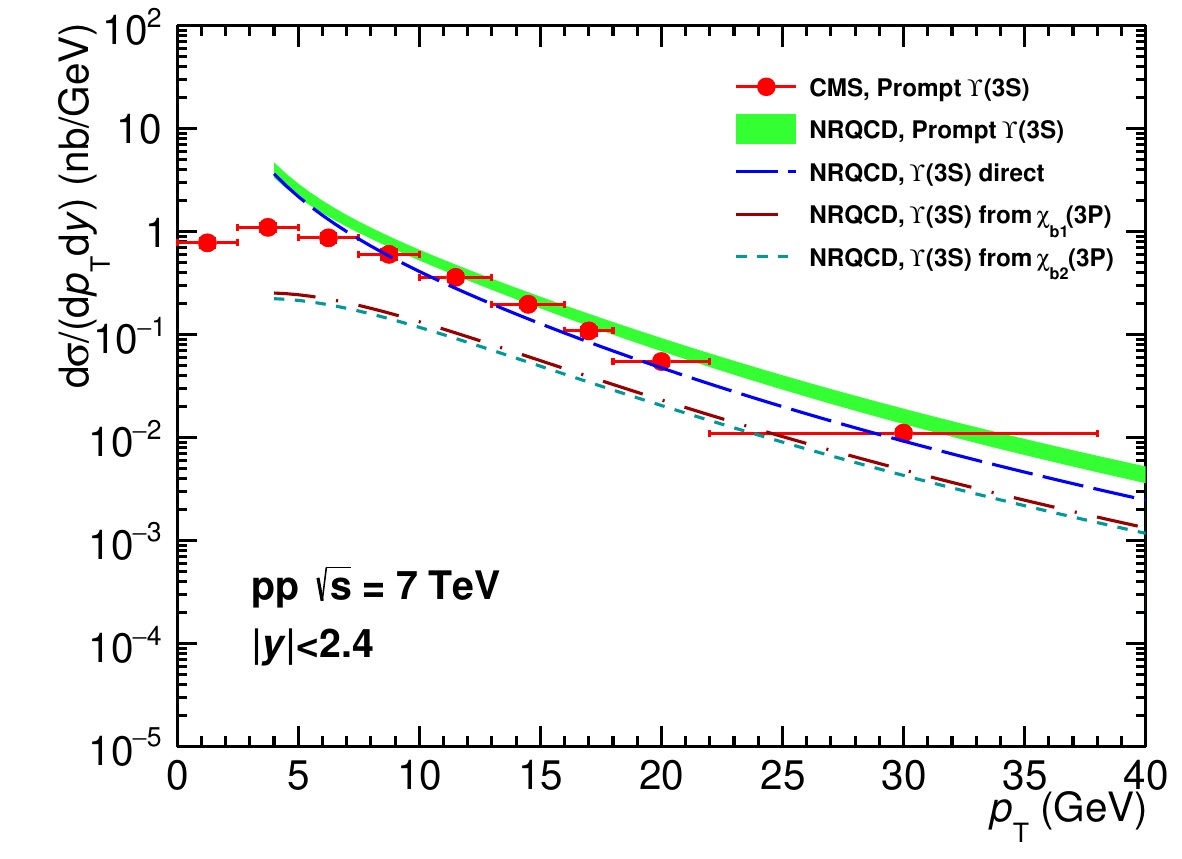}~~~~ \includegraphics[width=8.5cm,height=6.5cm,angle=0]{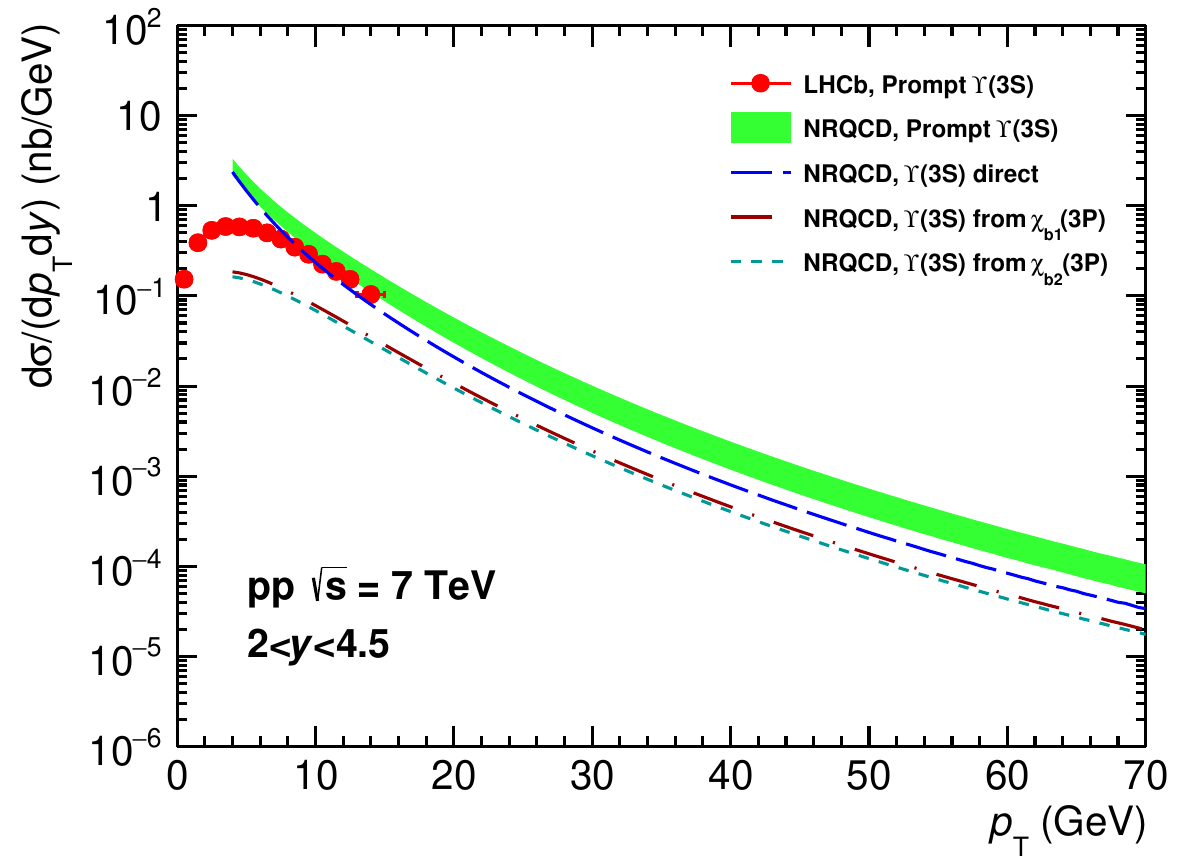}
 \caption{Differential production cross-section of $\Upsilon(3S)$ as a function of $p_{T}$ compared with the measurements by ATLAS~\cite{ATLAS_Upsilon_pp_7TeV}, CMS~\cite{CMS_Upsilon_pp_7TeV}, LHCb~\cite{LHCb_Upsilon_pp_7TeV} in p-p collisions at $\sqrt{s}$ = 7 TeV.}
 \label{Y3S_NRQCD}
 \end{figure*}

Figure~\ref{Y3S_NRQCD} presents the differential production cross-section of $\Upsilon(3S)$ as a function of $p_{\rm T}$, as obtained from our NRQCD calculations, together with the corresponding experimental data from ATLAS~\cite{ATLAS_Upsilon_pp_7TeV}, CMS~\cite{CMS_Upsilon_pp_7TeV}, and LHCb~\cite{LHCb_Upsilon_pp_7TeV}. The rapidity and $p_{\rm T}$ coverage for the $\Upsilon(3S)$ measurements in ATLAS and LHCb are identical to those for $\Upsilon(1S)$ and $\Upsilon(2S)$, while the CMS measurement extends up to $p_{\rm T}\approx38$ GeV in the mid-rapidity region ($|y|<2.4$). 
The NRQCD predictions corresponding to $\Upsilon(3S)$ production are shown as a green uncertainty band in Fig.~\ref{Y3S_NRQCD}. Overall, the calculated results provide a good description of the experimental measurements from all three Collaborations, particularly for $p_{\rm T} > 4$ GeV.

\begin{figure*}
    \centering
 \includegraphics[width=8.5cm,height=6.5cm,angle=0]{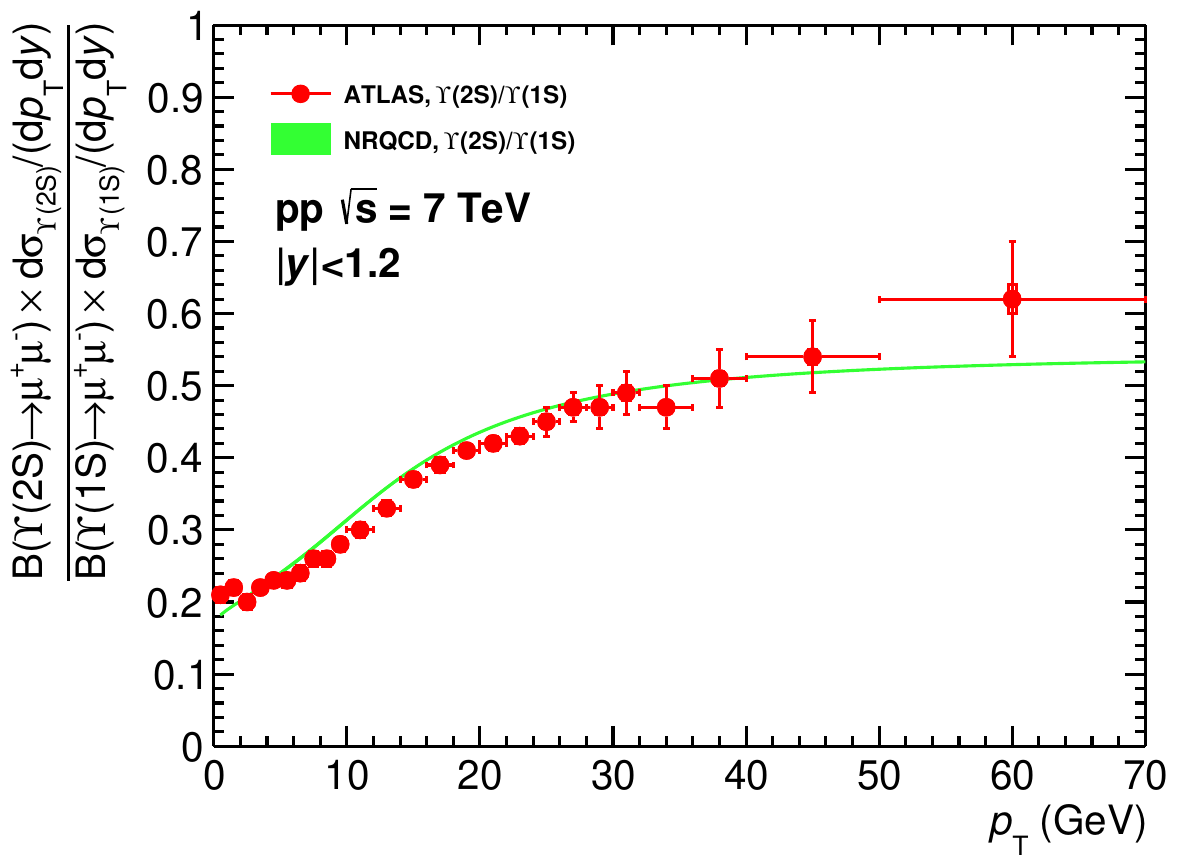}~~~~ \includegraphics[width=8.5cm,height=6.5cm,angle=0]{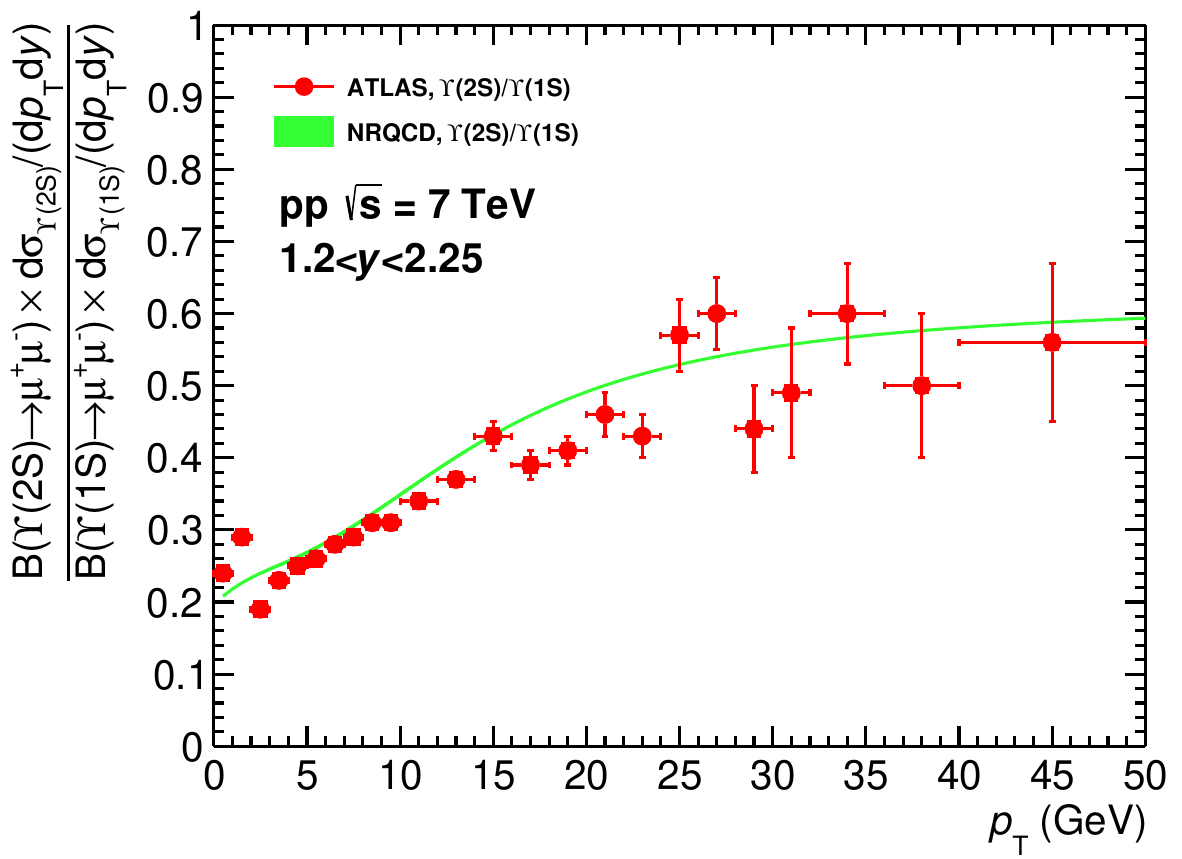}~~~~\\
  \includegraphics[width=8.5cm,height=6.5cm,angle=0]{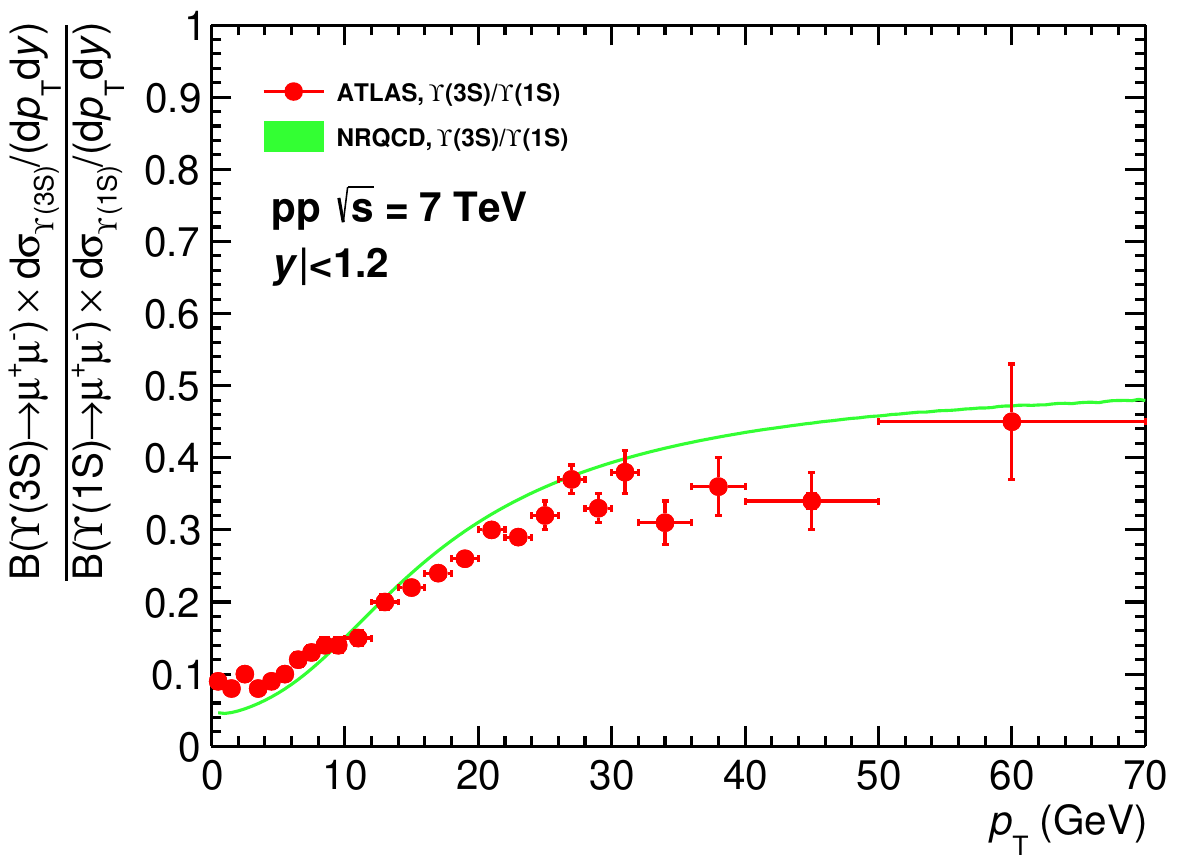}~~~~ \includegraphics[width=8.5cm,height=6.5cm,angle=0]{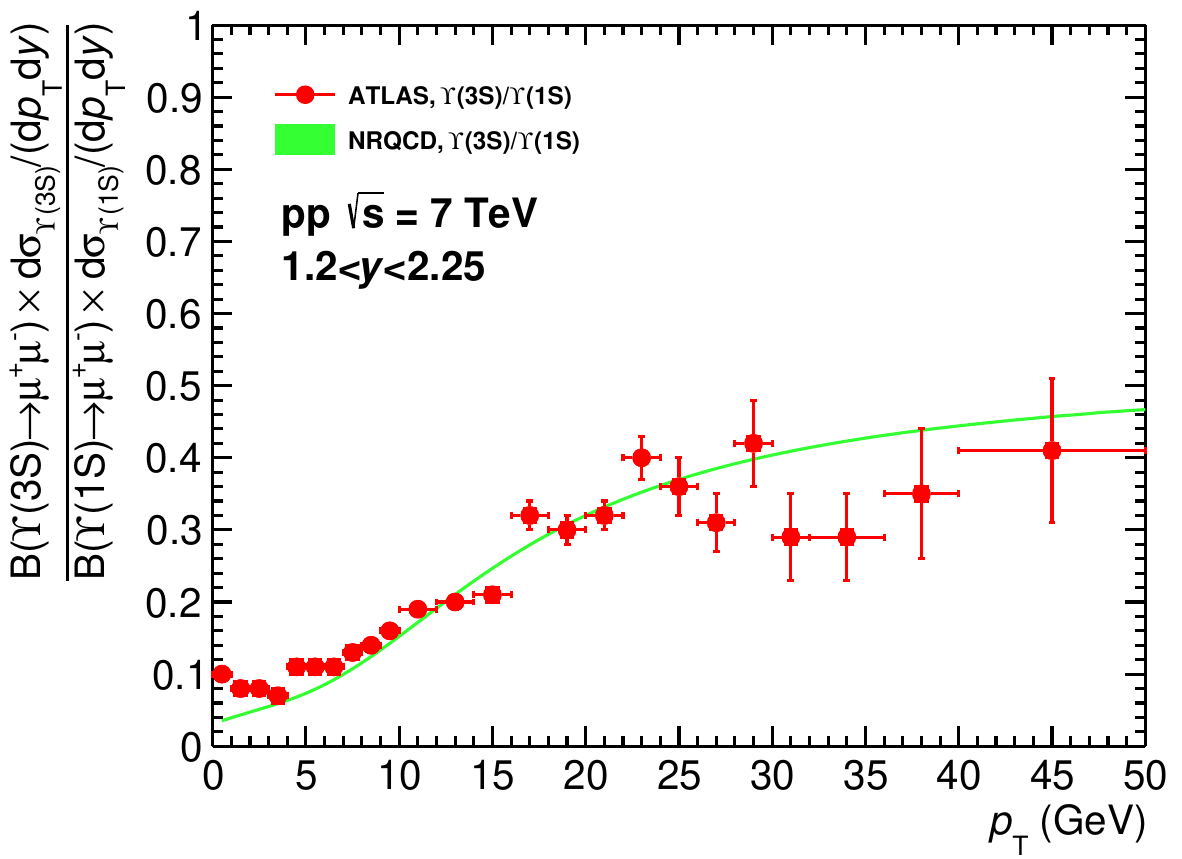}~~~~
 \caption{(Color online) $\Upsilon(2S)/\Upsilon(1S)$ production cross-section ratio (top panel) and $\Upsilon(3S)/\Upsilon(1S)$ production cross-section ratio (bottom panel) as a function of $p_{\rm T}$ compared with the measurements by ATLAS~\cite{ATLAS_Upsilon_pp_7TeV} in $pp$ collisions at $\sqrt{s}$ = 7 TeV.}
 \label{ATLAS_YnS_Y1S_NRQCD}
 \end{figure*}

\begin{figure*}
    \centering
 \includegraphics[width=8.5cm,height=6.5cm,angle=0]{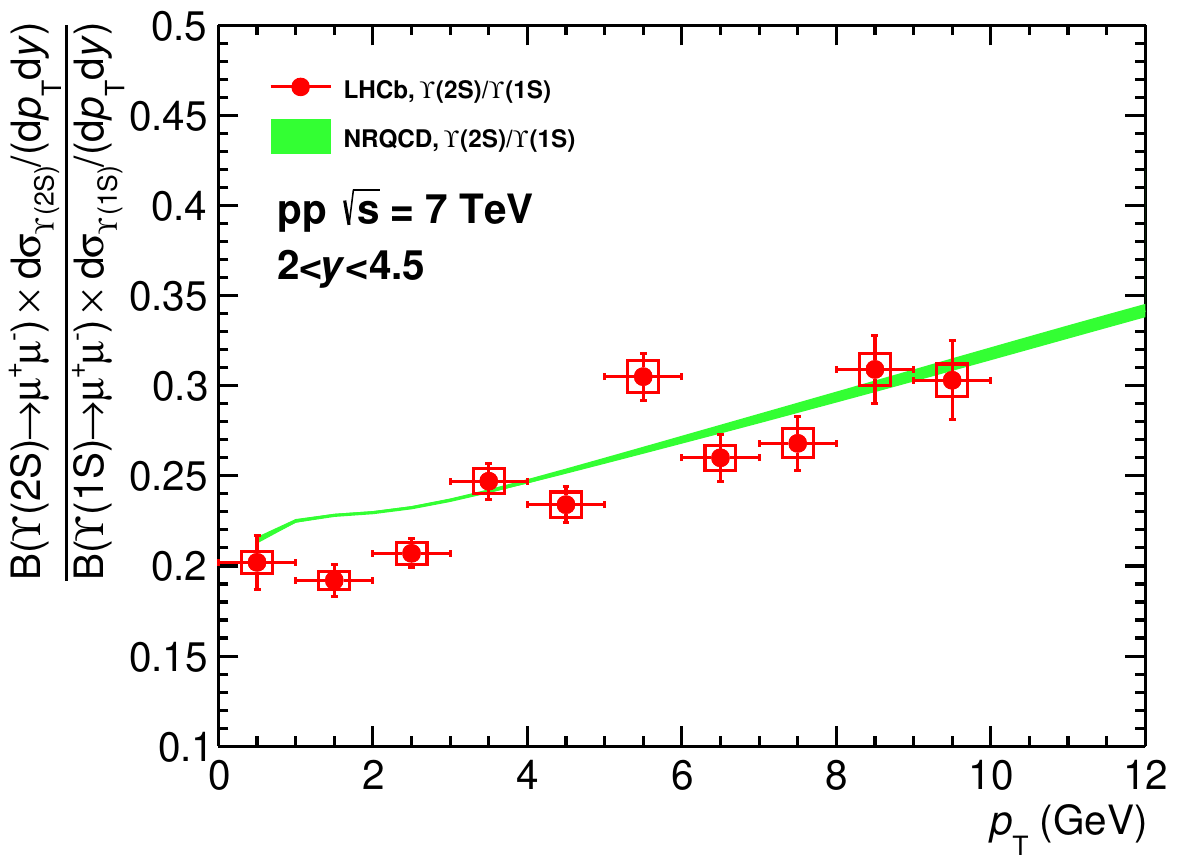}~~~~ \includegraphics[width=8.5cm,height=6.5cm,angle=0]{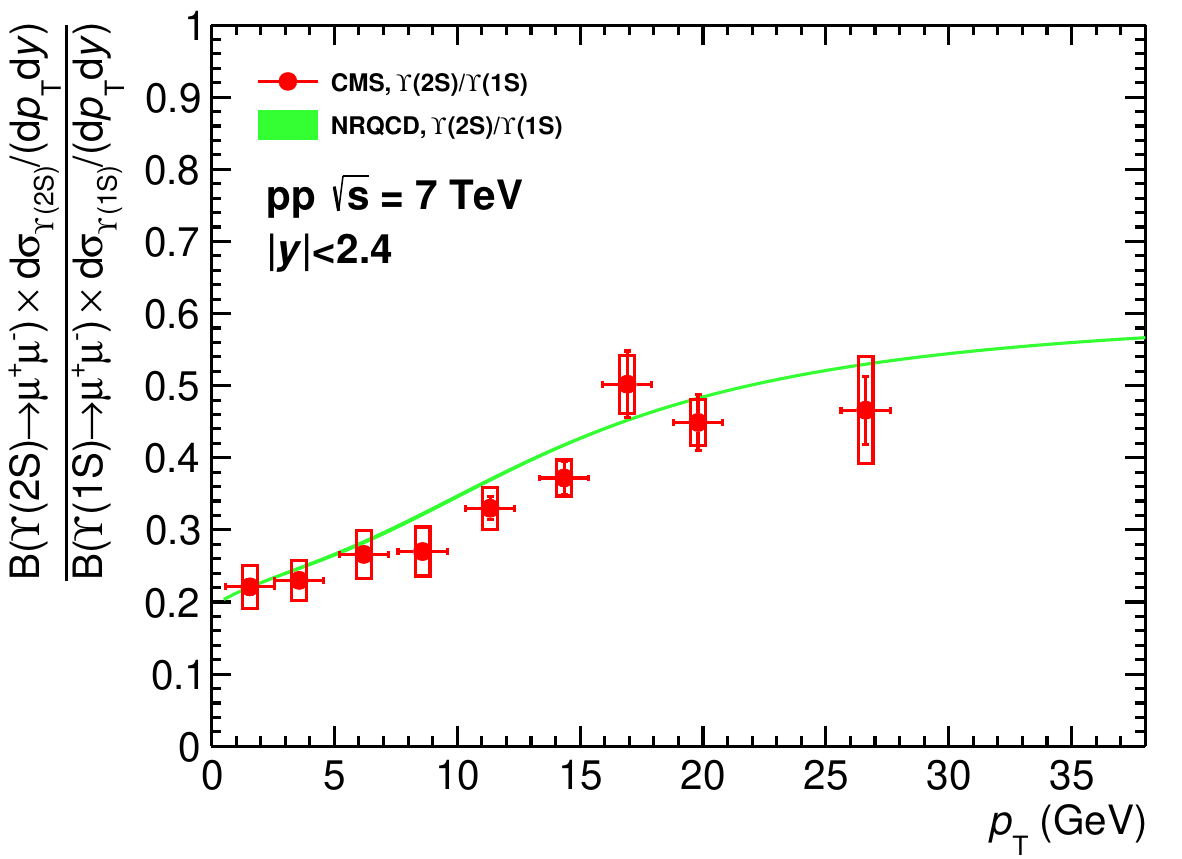}~~~~\\
  \includegraphics[width=8.5cm,height=6.5cm,angle=0]{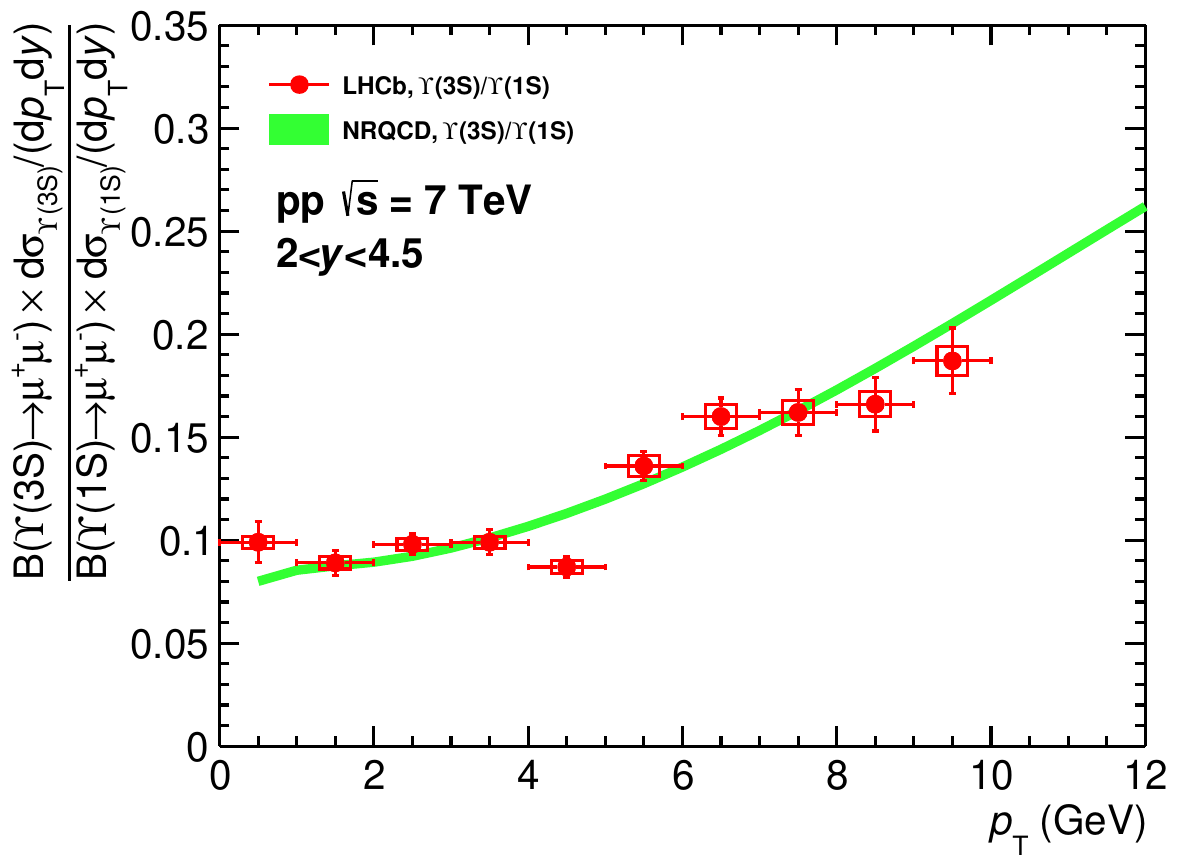}~~~~ \includegraphics[width=8.5cm,height=6.5cm,angle=0]{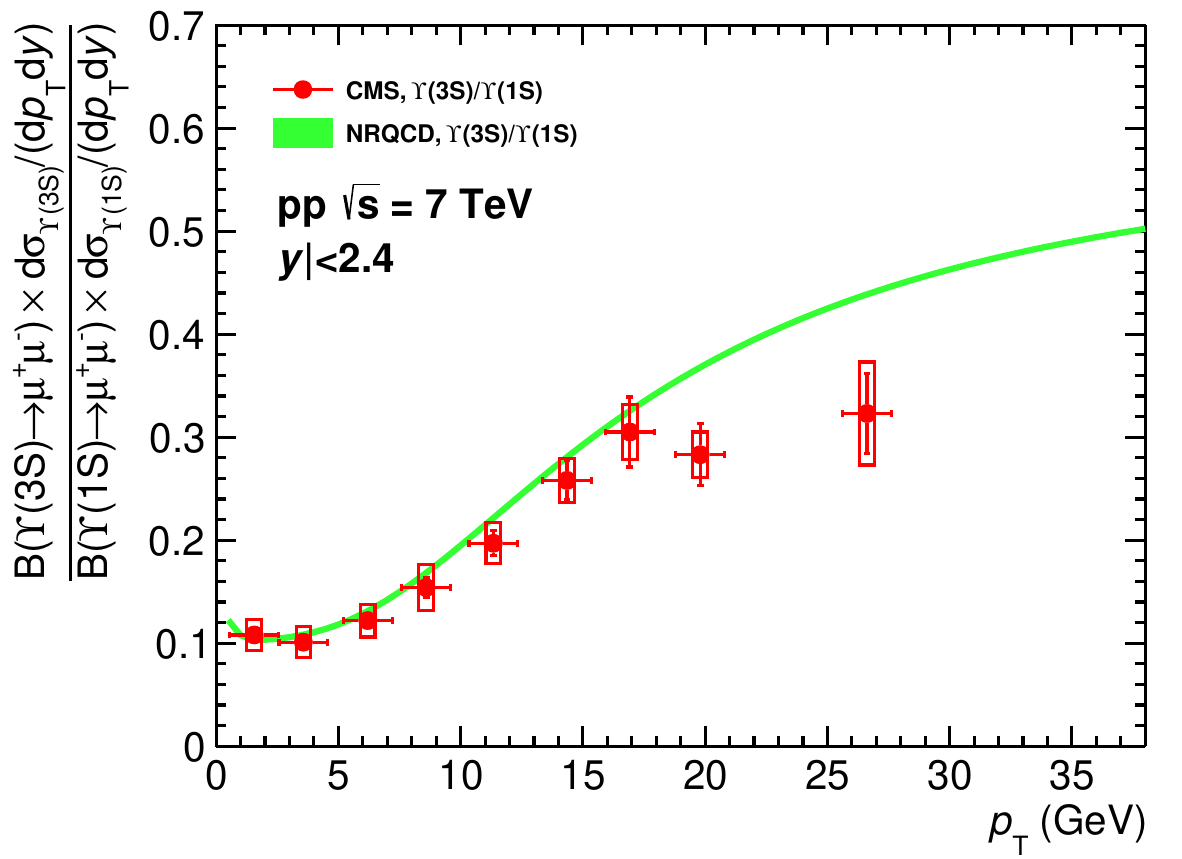}~~~~\\
    \includegraphics[width=8.5cm,height=6.5cm,angle=0]{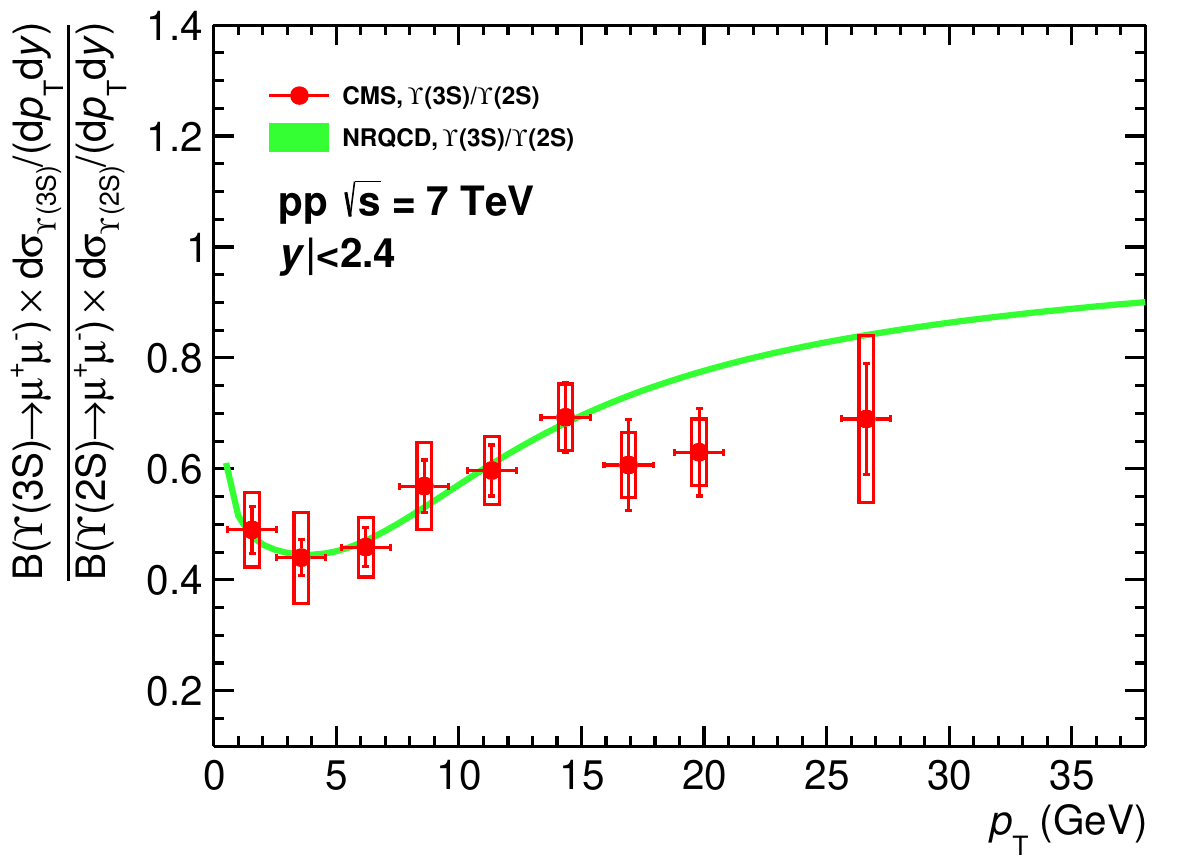}~~~~ 
 \caption{(Color online) $\Upsilon(2S)/\Upsilon(1S)$ production cross-section ratio (top panel), $\Upsilon(3S)/\Upsilon(1S)$ production cross-section ratio (middle panel) and $\Upsilon(3S)/\Upsilon(2S)$ production cross-section ratio (bottom panel) as a function of $p_{T}$ compared with the measurements by LHCb~\cite{LHCb_Upsilon_pp_7TeV} and CMS~\cite{CMS_Upsilon_pp_7TeV} in $pp$ collisions at $\sqrt{s}$ = 7 TeV.}
 \label{CMS_LHCb_YnS_Y1S_NRQCD}
 \end{figure*}

The $\Upsilon(2S)/\Upsilon(1S)$ and $\Upsilon(3S)/\Upsilon(1S)$ production cross-section ratios in $pp$ collisions at $\sqrt{s}=7$ TeV have been reported by ATLAS~\cite{ATLAS_Upsilon_pp_7TeV}, CMS~\cite{CMS_Upsilon_pp_7TeV} and LHCb~\cite{LHCb_Upsilon_pp_7TeV}. In addition, CMS has measured the $\Upsilon(3S)/\Upsilon(2S)$ cross-section ratio in the same collision system~\cite{CMS_Upsilon_pp_7TeV}. Our calculated ratios are compared with these experimental results in Fig.~\ref{ATLAS_YnS_Y1S_NRQCD} and Fig.~\ref{CMS_LHCb_YnS_Y1S_NRQCD}.
A reasonable agreement is observed across the entire $p_{\rm T}$ range ($p_{\rm T} > 0$) for all experiments covering mid-, near-forward-, and forward-rapidity intervals, despite the fact that the individual cross-sections are best reproduced for $p_{\rm T} > 4$ GeV. The increasing trend of these ratios with $p_{\rm T}$ is well captured by our NRQCD calculations. 

Furthermore, as seen in Fig.~\ref{ATLAS_YnS_Y1S_NRQCD}, our results exhibit a saturation behavior of the cross-section ratios beyond $p_{\rm T}\approx40$ GeV, indicating a nearly flat dependence at high $p_{\rm T}$. The ATLAS data also show a similar trend within experimental uncertainties. It will be of significant interest to investigate, with future high-luminosity measurements, whether this saturation persists at even higher $p_{\rm T}$ values.

\begin{figure*}
    \centering
 \includegraphics[width=8.5cm,height=6.5cm,angle=0]{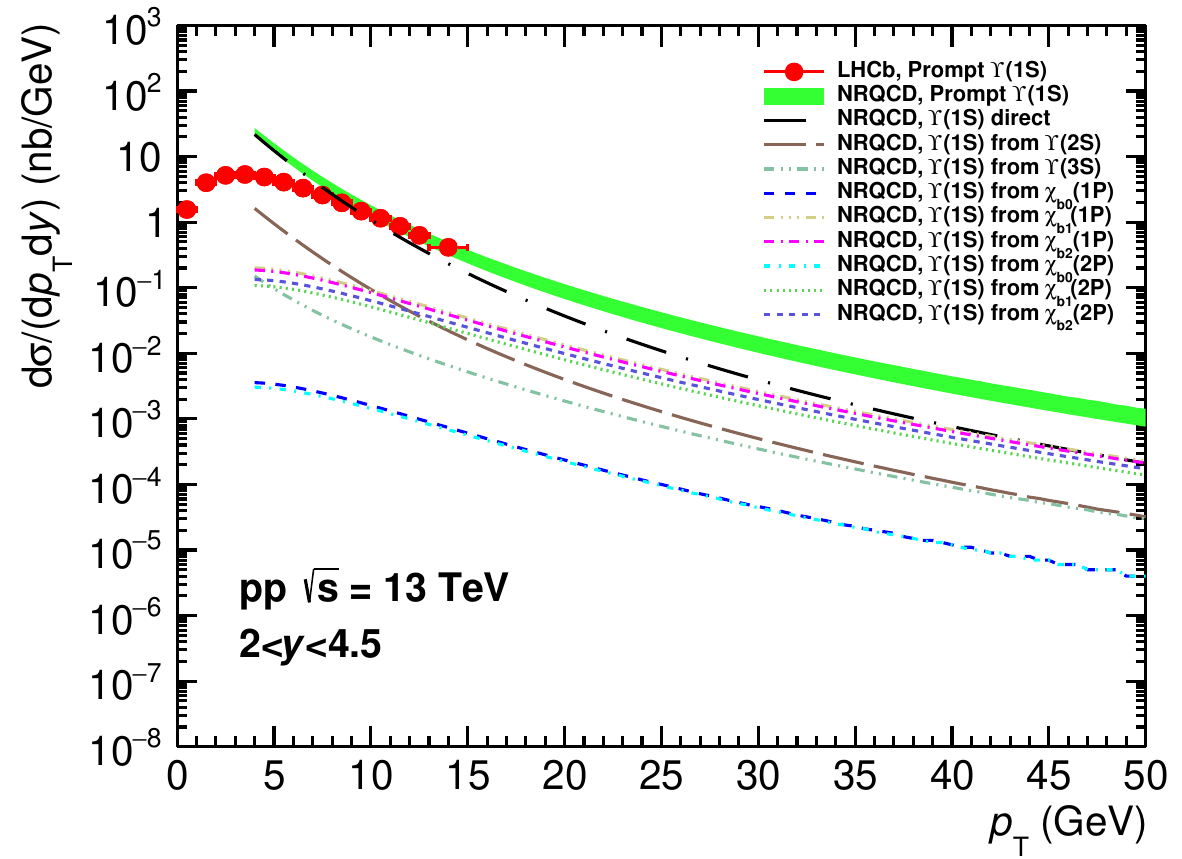}~~~~ \includegraphics[width=8.5cm,height=6.5cm,angle=0]{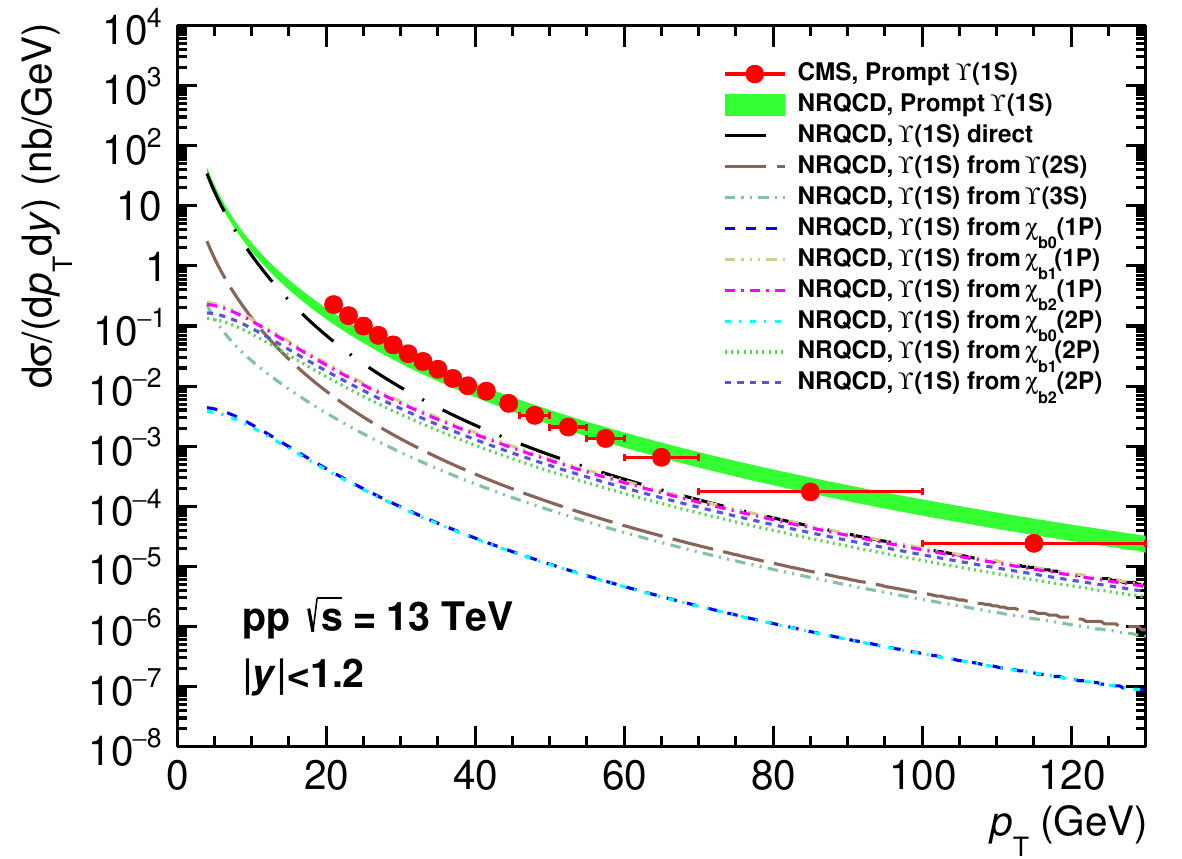}~~~~~\\  \includegraphics[width=8.5cm,height=6.5cm,angle=0]{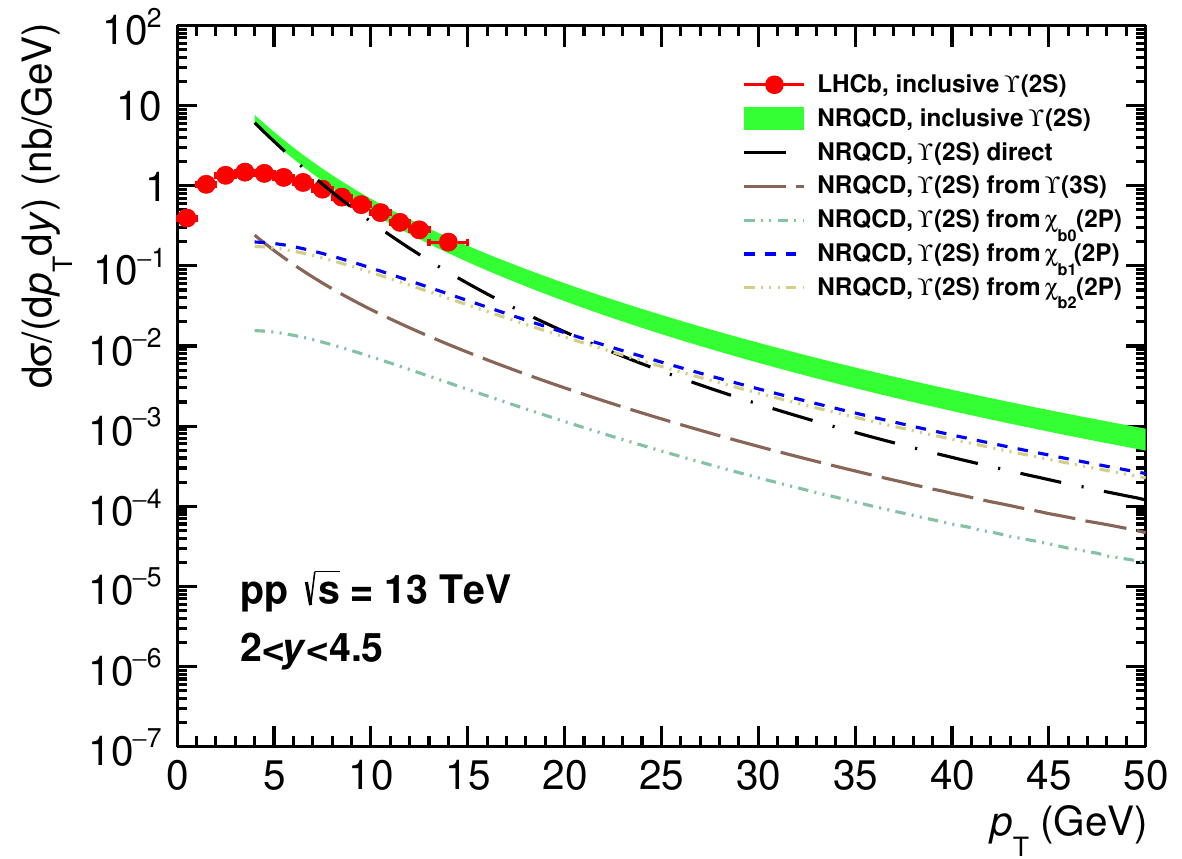}~~~~ \includegraphics[width=8.5cm,height=6.5cm,angle=0]{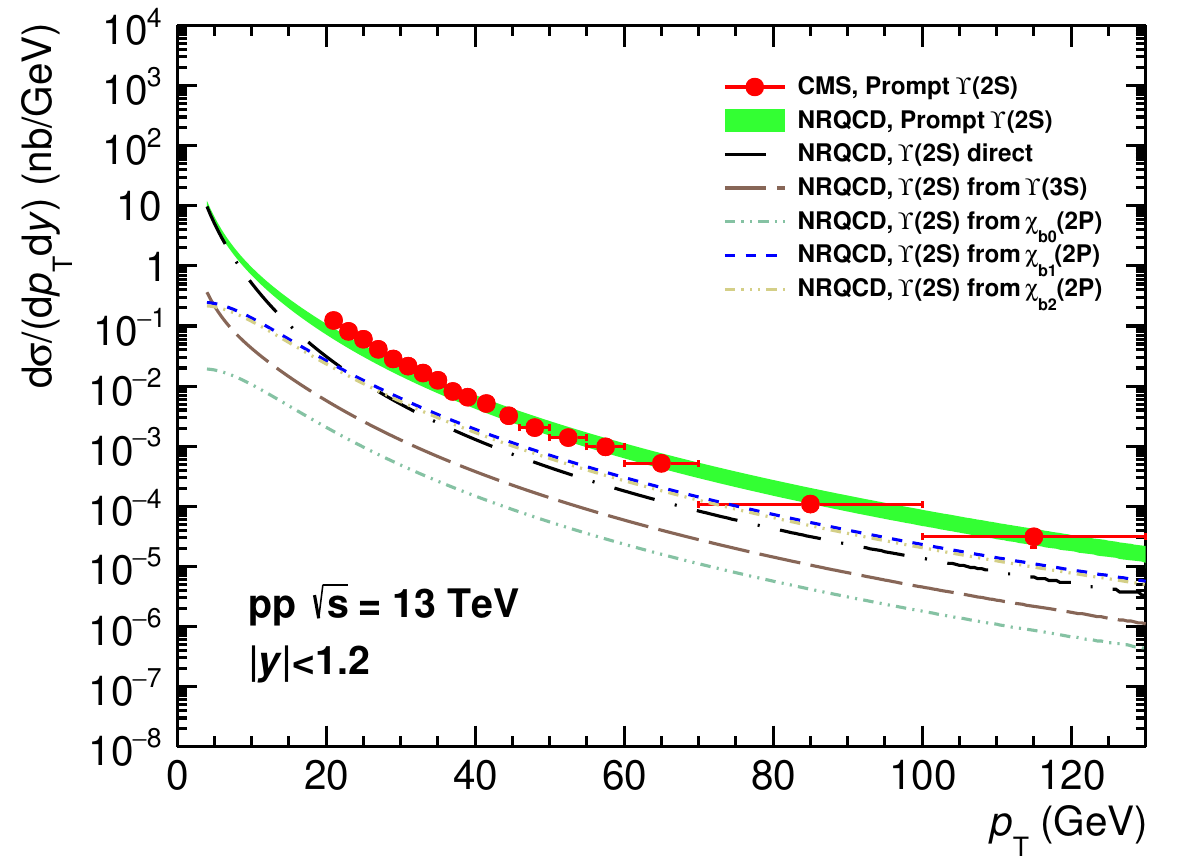}~~~~~~~\\
 \includegraphics[width=8.5cm,height=6.5cm,angle=0]{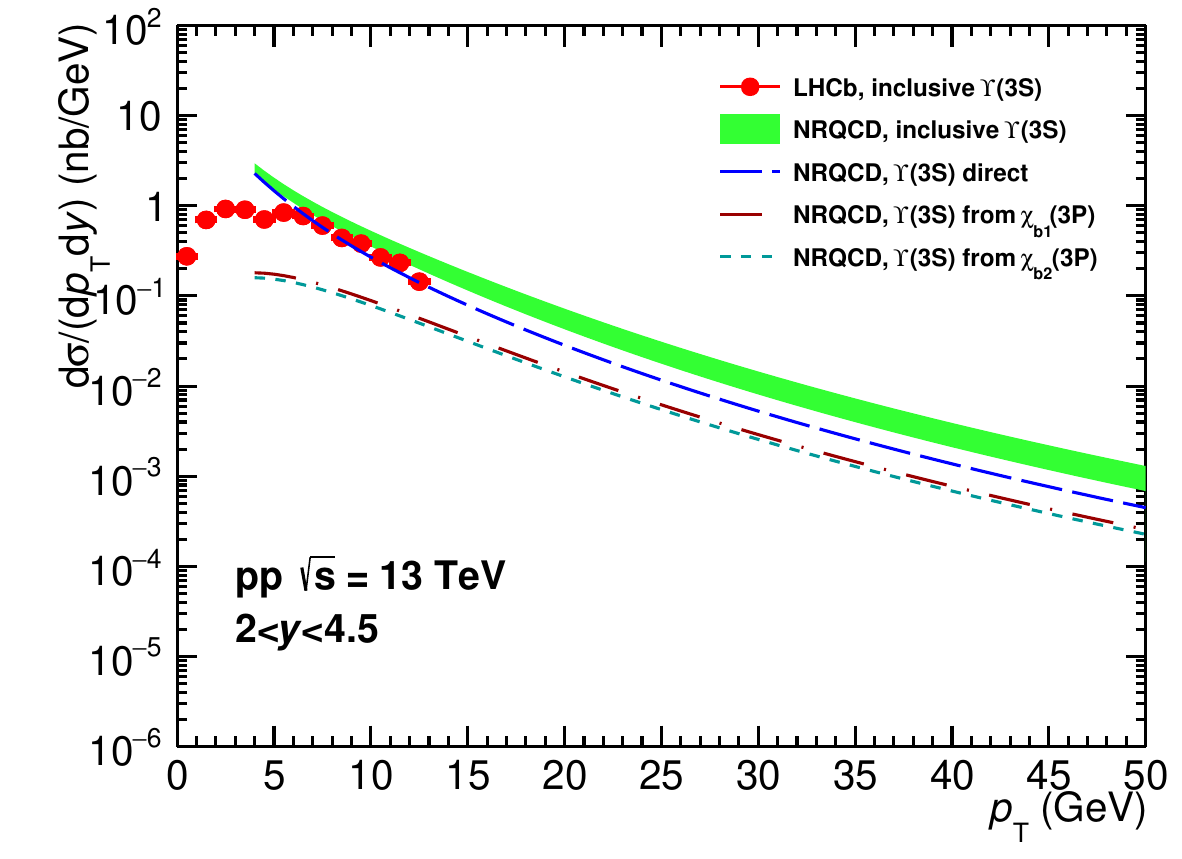}~~~~ \includegraphics[width=8.5cm,height=6.5cm,angle=0]{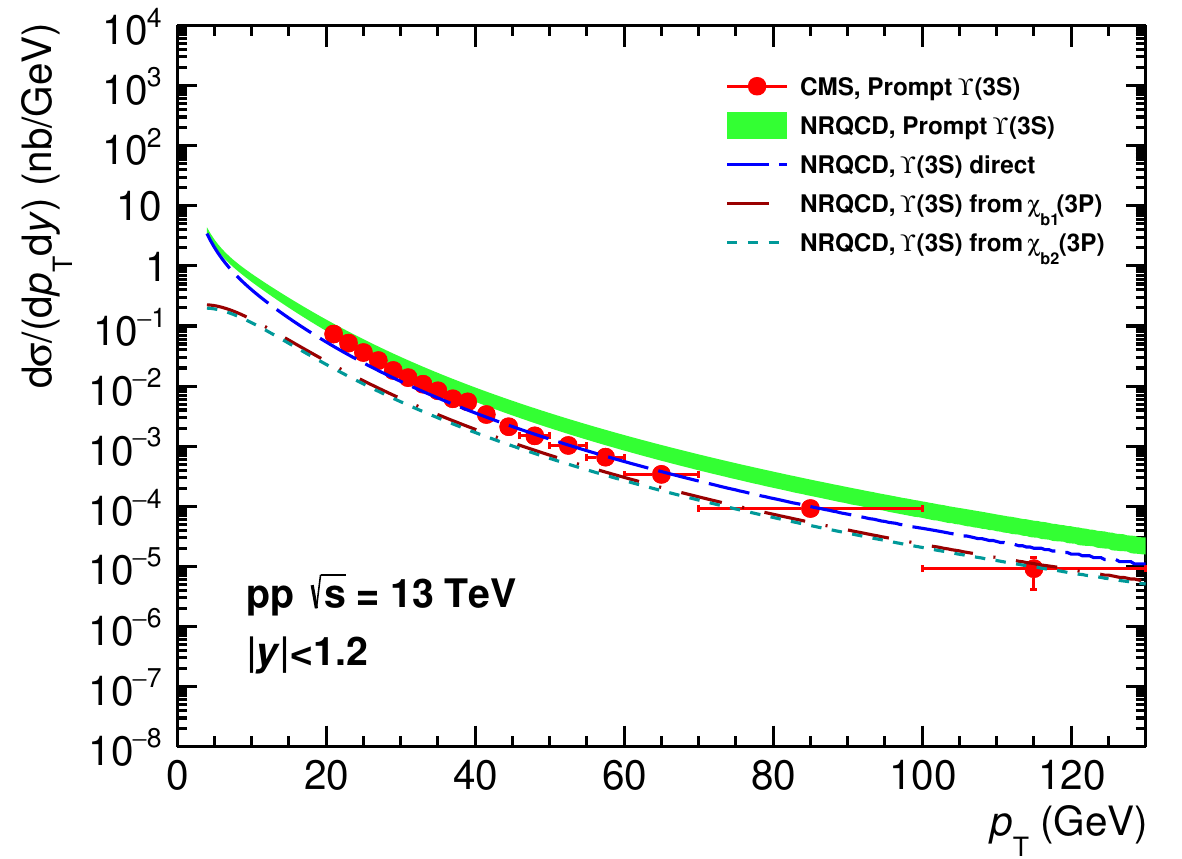}~~~~~
 \caption{Differential production cross-section of $\Upsilon(1S)$ (top), $\Upsilon(2S)$ (middle) and $\Upsilon(3S)$ (bottom) as a function of $p_{T}$ compared with the measurements by LHCb~\cite{LHCb_Upsilon_pp_7TeV} and CMS~\cite{CMS_Upsilon_pp_13TeV} in $pp$ collisions at $\sqrt{s}$ = 13 TeV.}
 \label{CMS_LHCb_YnS_13TeV_NRQCD}
 \end{figure*}

Figure~\ref{CMS_LHCb_YnS_13TeV_NRQCD} presents the differential cross-sections of $\Upsilon(nS)$ as a function of $p_{\rm T}$ obtained from our NRQCD calculations, compared with the experimental data reported by LHCb~\cite{LHCb_Upsilon_pp_13TeV} and CMS~\cite{CMS_Upsilon_pp_13TeV} in $pp$ collisions at $\sqrt{s}$ = 13 TeV.
LHCb has measured the $\Upsilon(nS)$ production in the forward rapidity region ($2 < y < 4.5$) over the $p_{\rm T}$ range 0 $<$ $p_{\rm T}$ $<$ 13 GeV, while CMS has performed measurements at midrapidity ($|y| < 1.2$) for 20 $<$ $p_{\rm T}$ $<$ 130 GeV. The CMS dataset extends up to 130 GeV in $p_{\rm T}$, representing the highest $p_{\rm T}$ reach for bottomonium production in $pp$ collisions at the LHC, thus providing an excellent opportunity to test the NRQCD framework in an extended kinematic regime. 
In the figure, the total NRQCD predictions, obtained by summing all contributions, are shown as a green band, while the direct and feed-down components from higher bottomonium states are represented by lines for their central values. The results clearly demonstrate that the feed-down contributions play a crucial role in shaping the overall $p_{\rm T}$ dependence of the prompt $\Upsilon(1S)$ and $\Upsilon(2S)$ cross-sections. Overall, our NRQCD calculations show good agreement with the experimental data, particularly for $p_{\rm T}$ $>$ 4 GeV.

\begin{figure*}
    \centering
 \includegraphics[width=8.5cm,height=6.5cm,angle=0]{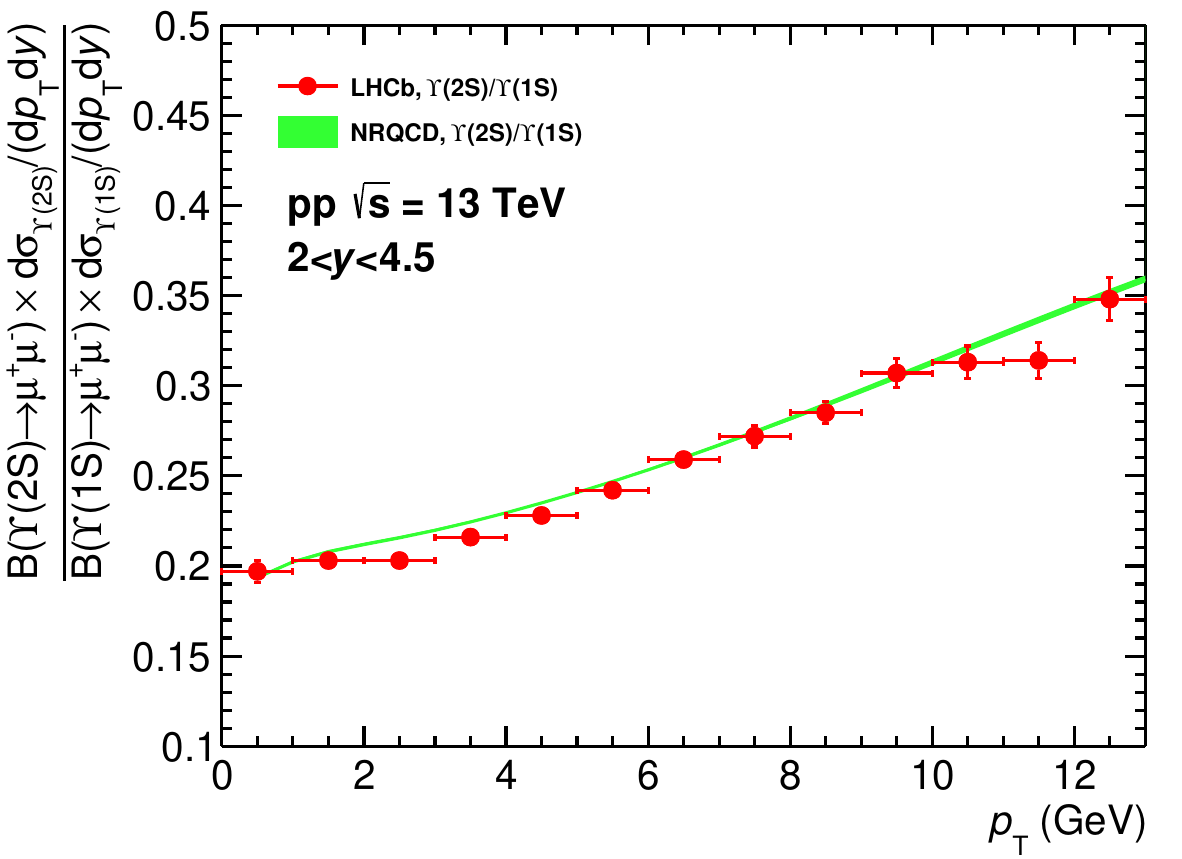}~~~~ \includegraphics[width=8.5cm,height=6.5cm,angle=0]{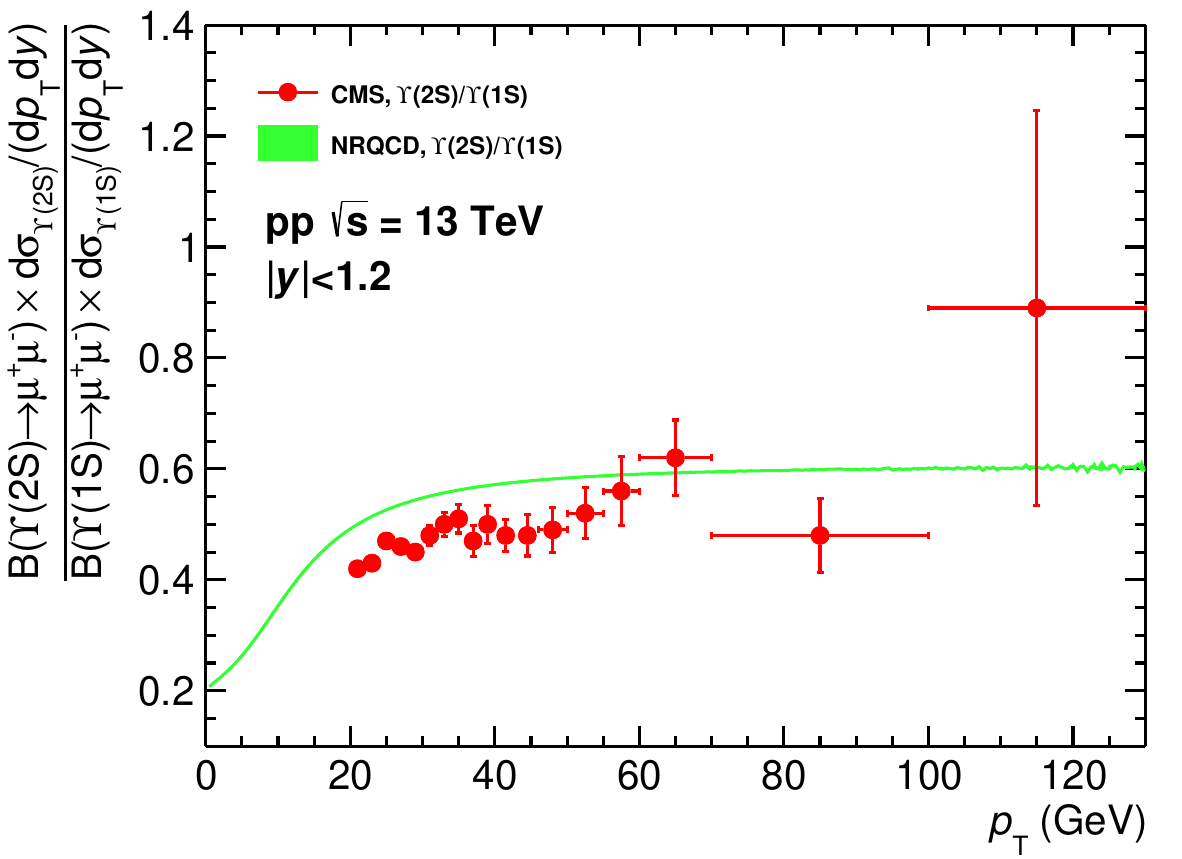}~~~~~\\
\includegraphics[width=8.5cm,height=6.5cm,angle=0]{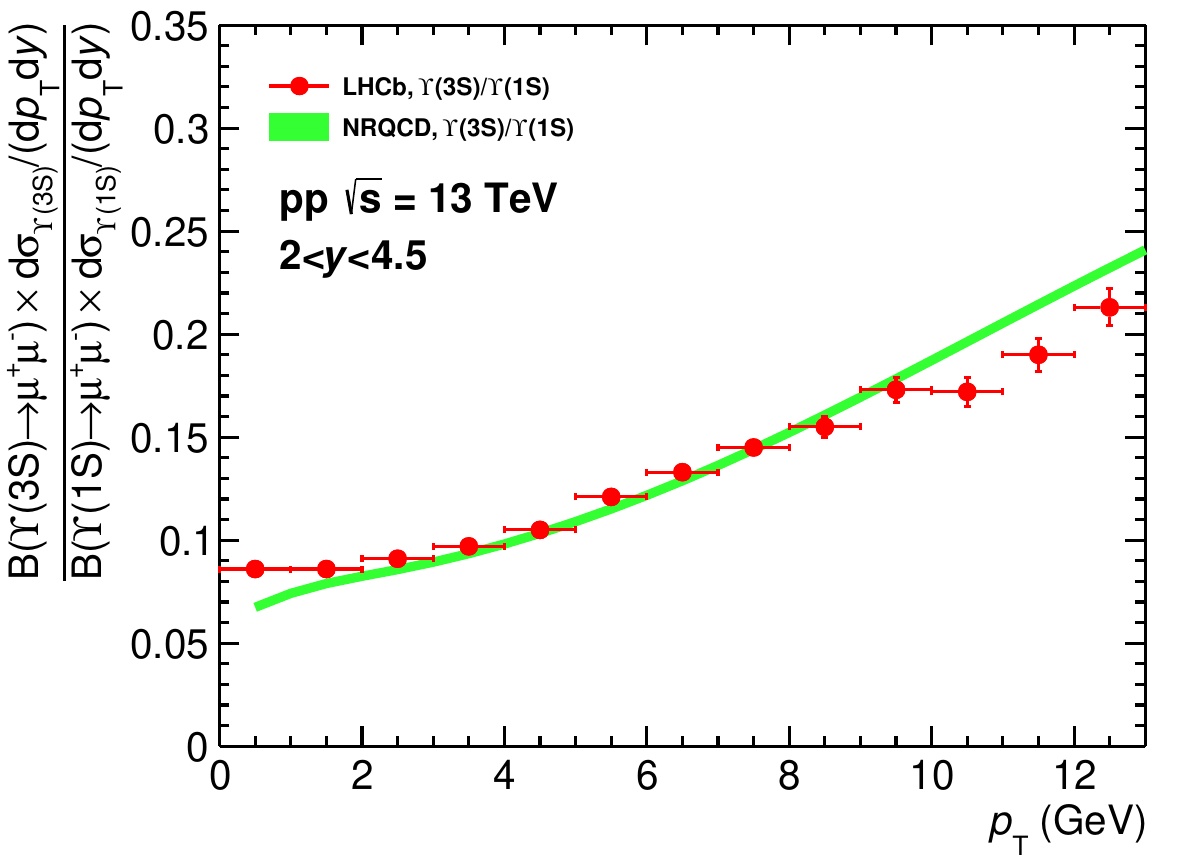}~~~~ \includegraphics[width=8.5cm,height=6.5cm,angle=0]{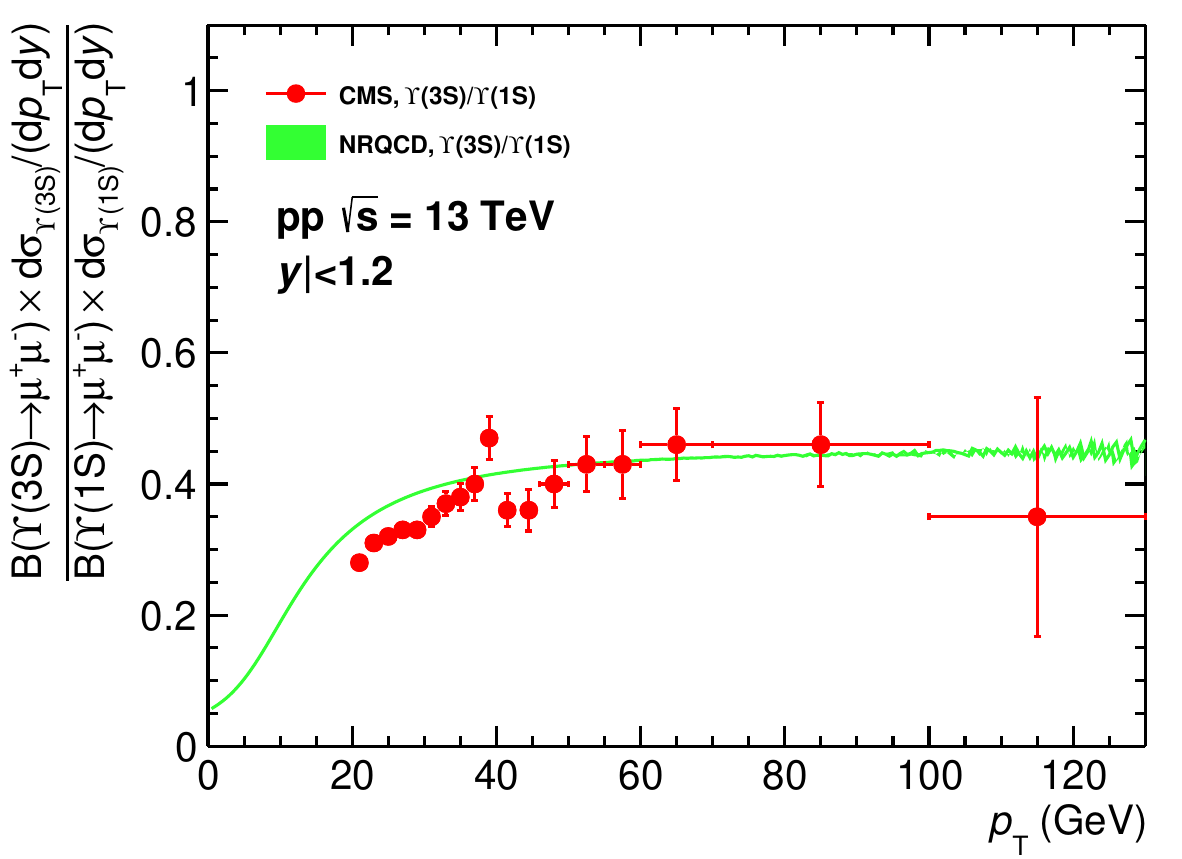}~~~~~~\\
\includegraphics[width=8.5cm,height=6.5cm,angle=0]{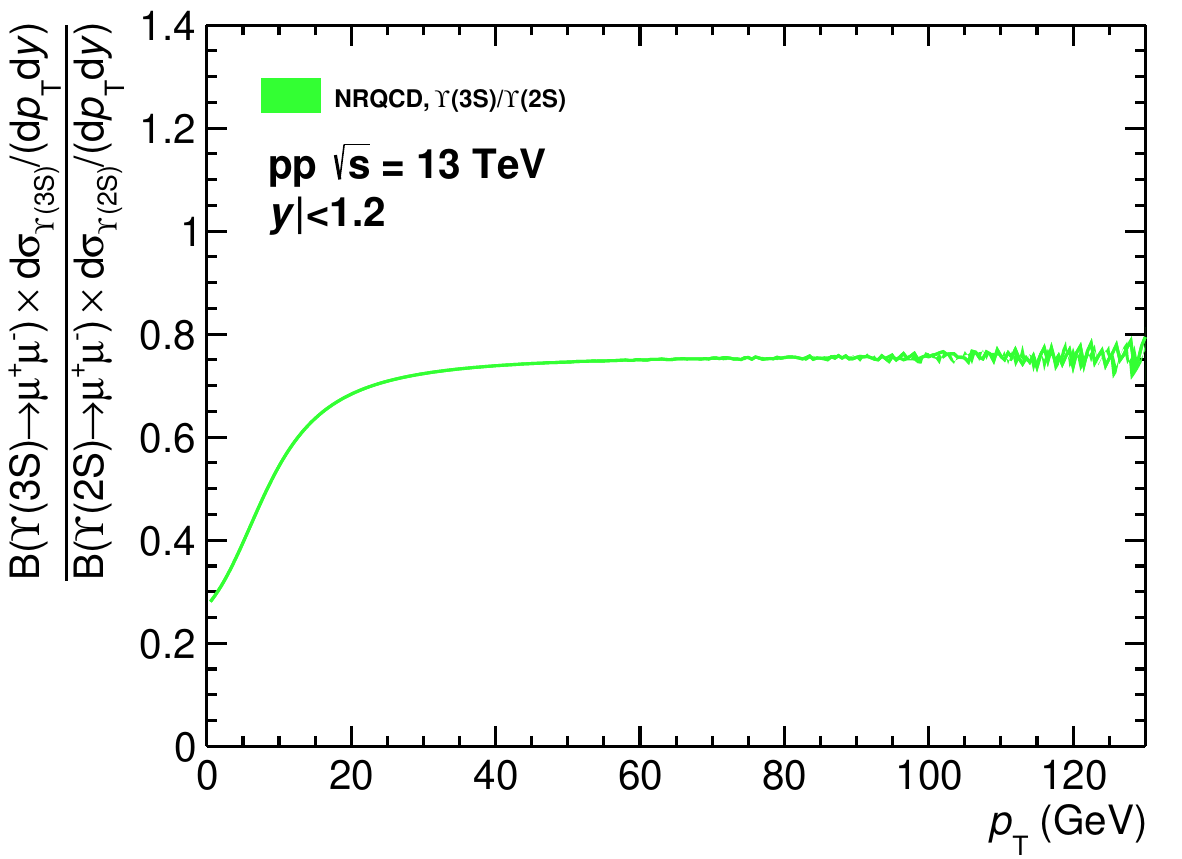}~~~~ 
 \caption{(Color online) $\Upsilon(2S)/\Upsilon(1S)$ production cross-section ratio (top) and $\Upsilon(3S)/\Upsilon(1S)$ production cross-section ratio (middle) as a function of $p_{T}$ compared with the measurements by LHCb~\cite{LHCb_Upsilon_pp_13TeV} and CMS~\cite{CMS_Upsilon_pp_13TeV} and prediction for $\Upsilon(3S)/\Upsilon(2S)$ production cross-section ratio (bottom) in $pp$ collisions at $\sqrt{s}$ = 13 TeV.}
 \label{CMS_LHCb_YnS_Y2S_13TeV_NRQCD}
 \end{figure*}

The $\Upsilon(2S)/\Upsilon(1S)$ and $\Upsilon(3S)/\Upsilon(1S)$ production cross-section ratios in $pp$ collisions at $\sqrt{s}=13$ TeV have been reported by the CMS~\cite{CMS_Upsilon_pp_13TeV} and LHCb~\cite{LHCb_Upsilon_pp_13TeV} Collaborations. Our NRQCD calculations are compared with these experimental results in Fig.~\ref{CMS_LHCb_YnS_Y2S_13TeV_NRQCD}. In addition, prediction for the $\Upsilon(3S)/\Upsilon(2S)$ cross-section ratio at midrapidity ($|y| < 1.2$) in the same collision system is also shown in Fig.~\ref{CMS_LHCb_YnS_Y2S_13TeV_NRQCD}. A consistent agreement is observed across the entire transverse momentum range ($p_{\rm T} > 0$) and for all rapidity regions, although as also noted at $\sqrt{s}=7$ TeV, the best description of the absolute cross-sections is achieved for $p_{\rm T} > 4$ GeV. The observed rise of the cross-section ratios with increasing $p_{\rm T}$ is well reproduced by our NRQCD calculations, confirming the robustness of the theoretical framework.

At higher transverse momenta, our results display a clear saturation behavior of the cross-section ratios beyond $p_{\rm T} \approx 40$ GeV, indicating a nearly constant dependence at large $p_{\rm T}$. The CMS data exhibit a comparable trend within experimental uncertainties up to the highest measured $p_{\rm T}$ value of about 130 GeV. This agreement strongly suggests that the saturation of the production ratios persists at even higher transverse momenta, reflecting the underlying scaling behavior of bottomonium production at LHC energies.

\section{Summary and discussion}
In this work, we have carried out a comprehensive study of $\Upsilon(nS)$ production in $pp$ collisions at various LHC energies and rapidity ranges within the framework of NRQCD. The calculations account for both direct production and feed-down contributions from higher bottomonium states. For $\Upsilon(1S)$ production, feed-down from $\Upsilon(2S)$, $\Upsilon(3S)$, $\chi_{bJ}(1P)$ and $\chi_{bJ}(2P)$ is included, while $\Upsilon(2S)$ receives contributions from $\Upsilon(3S)$ and $\chi_{bJ}(2P)$. For $\Upsilon(3S)$, feed-down from $\chi_{bJ}(3P)$ states is also incorporated, providing a consistent treatment of the prompt $\Upsilon(nS)$ yields.


Our results demonstrate that the calculated $p_{\rm T}$ dependent cross-sections of $\Upsilon(nS)$ are in good agreement with the experimental measurements from ATLAS, CMS, LHCb, and ALICE over a wide range of $p_{\rm T}$ and rapidity. The inclusion of feed-down contributions was found to be crucial for accurately reproducing the observed $p_{\rm T}$-dependent shapes. The model reproduces the data well for $p_{\rm T} > 4$ GeV. 

We have also studied the production cross-section ratios $\Upsilon(2S)/\Upsilon(1S)$, $\Upsilon(3S)/\Upsilon(1S)$ and $\Upsilon(3S)/\Upsilon(2S)$ at $\sqrt{s}$ = 7 and 13 TeV and compared them with the ATLAS, CMS, LHCb and ALICE results. The NRQCD calculations successfully capture the increasing trend of these ratios with $p_{\rm T}$, followed by a saturation behavior beyond $p_{\rm T} \approx 40$ GeV. This flattening of the ratios at high $p_{\rm T}$—also observed in the CMS data up to $p_{\rm T} \approx 130$ GeV—suggests a universal scaling feature of bottomonium production mechanisms at large $p_{\rm T}$. 

The saturation of cross-section ratios at high $p_{\mathrm{T}}$ arises because bottomonium production enters the fragmentation-dominated regime, where the short-distance partonic cross-sections become universal across states, and the only differences are encoded in constant nonperturbative matrix elements. This reflects the asymptotic scaling behavior predicted by NRQCD and confirms the factorization of short- and long-distance physics in heavy-quarkonium production.

The present analysis also emphasizes the need for improved experimental
information on the $3P$ bottomonium states. In particular, precise experimental measurements of the $\chi_{bJ}(3P)\to\Upsilon(3S)$ branching fractions and
the mass of the $\chi_{b0}(3P)$ state are needed to better constrain the
feed-down contribution to $\Upsilon(3S)$ production.

Overall, the present results provide a consistent phenomenological description
of $\Upsilon(nS)$ production and their ratios at the LHC within the LO NRQCD
framework. The study emphasizes the importance of treating direct production
and feed-down contributions consistently when comparing different
bottomonium states. Future improvements should include higher-order QCD
corrections, more precise information on the $3P$ bottomonium states, and
polarization observables. Extending the analysis to heavy-ion collisions would
also provide an opportunity to investigate bottomonium production in the
presence of cold and hot nuclear-matter effects.



\section*{Acknowledgements}
It is a pleasure to thank Mahatsab Mandal, Pradip Roy and
Sukalyan Chattopadhyay for helpful discussions.



\end{document}